\documentclass[12pt]{article}
\usepackage{epsf}

\begin{document}

\newcommand{\gtrsim}{ \mathop{}_{\textstyle \sim}^{\textstyle >} }
\newcommand{\lesssim}{ \mathop{}_{\textstyle \sim}^{\textstyle <} }

\newcommand{\rem}[1]{{\bf #1}}

\renewcommand{\thefootnote}{\fnsymbol{footnote}}
\setcounter{footnote}{0}
\begin{titlepage}

\def\thefootnote{\fnsymbol{footnote}}

\begin{center}

\hfill TU-755\\
\hfill September, 2005\\

\vskip .75in

{\Large \bf 

Supersymmetric Heavy Higgses 
at $e^+e^-$ Linear Collider
and Dark-Matter Physics

}

\vskip .75in

{\large
Takeo Moroi and Yasuhiro Shimizu
}

\vskip 0.25in

{\em
Department of Physics, Tohoku University,
Sendai 980-8578, JAPAN}

\end{center}

\vskip .5in

\begin{abstract}

We consider the capability of the $e^+e^-$ linear collider (which is
recently called as the International Linear Collider, or ILC) for
studying the properties of the heavy Higgs bosons in the
supersymmetric standard model.  We pay special attention to the
large $\tan\beta$ region which is motivated, in particular, by
explaining the dark-matter density of the universe (i.e., so-called
``rapid-annihilation funnels'').  We perform a systematic analysis to
estimate expected uncertainties in the masses and
widths of the heavy Higgs bosons 
assuming an energy and integrated luminosity of 
$\sqrt{s}=1$ TeV and $L=1$ ab$^{-1}$.
We also discuss its implication
to the reconstruction of the dark-matter density of the universe.

\end{abstract}

\end{titlepage}

\renewcommand{\thepage}{\arabic{page}}
\setcounter{page}{1}
\renewcommand{\thefootnote}{\#\arabic{footnote}}
\setcounter{footnote}{0}
\renewcommand{\theequation}{\thesection.\arabic{equation}}

\section{Introduction}
\setcounter{equation}{0}
\label{sec:intro}

An $e^+e^-$ linear collider, the
International Linear Collier (ILC), has been discussed as a new
experimental tool to study the physics at the electroweak scale and
beyond \cite{JLC1,NLC,Aguilar-Saavedra:2001rg,Snowmass}.  The primary
purpose of the ILC is to clarify the physics of the electroweak
symmetry breaking and physics beyond the standard model; in order to
solve the naturalness problem in the standard model, it is expected
that some new physics shows up at the energy scale of $100\ {\rm GeV}$
$-$ $1\ {\rm TeV}$.  Discovery and, in particular, precise studies of
such new physics should become very important once the ILC will be
built.  Thus, at the current stage, we should study the potential of
the ILC for such purposes.

Among various possibilities, supersymmetry (SUSY) is one of the
prominent candidates of the new physics beyond the standard model.
Accordingly, superpartners of the standard-model particles are
regarded as significant targets of the ILC.  As well as the
superparticles, however, we should also study another class of new
particles which show up in the supersymmetric models, that is, the
heavy Higgs bosons.  Since the minimal supersymmetric standard model
(MSSM) contains two Higgs doublets, i.e., up- and down-type Higgs
bosons, CP even and odd neutral Higgs bosons ($H$ and $A$) as well as
the charged Higgses $H^\pm$ exist in the physical spectrum as well as
the standard-model-like Higgs $h$.  Detailed study of the heavy
Higgses provides information about some of the parameters in the MSSM
like, for example, $\tan\beta$
\cite{Djouadi:1996pj,Gunion:1996cc,Feng:1996xv,Barger:2000fi,
Gunion:2002ip, Desch:2004yb}.  
Thus, the study of the heavy Higgses will be
important for the study of the Higgs potential and electroweak
symmetry breaking.

The study of the heavey Higgses is important both from the point of view
of understanding electroweak symmetry breaking as well as for its deep 
impacts on cosmology.
It has long been recognized that the dark matter in the
universe, whose origin cannot be explained in the framework of the
minimal standard model, may be explained in the MSSM.  With the
$R$-parity conservation, the lightest superparticle (LSP) is stable
and hence, if its interaction is weak enough, LSP can be a viable
candidate for the cold dark matter.  Importantly, in large fraction of
the parameter space, the lightest neutralino becomes the LSP which can
be the cold dark matter in this case.  In the past, the relic density
of the lightest neutralino has been intensively studied
\cite{Baer:2002gm,Ellis:2003cw,Baer:2003yh,Chattopadhyay:2003xi,
Lahanas:2003yz,Baer:2003wx,Battaglia:2003ab,Belanger:2004ag,
Baltz:2004aw}.

It should be noticed that, in order to realize the LSP dark matter,
there is one caveat; in many cases, the thermal relic density of the
LSP becomes larger than the currently measured dark matter density.
In particular, the density parameter of the dark matter is now well
constrained by the WMAP \cite{Bennett:2003bz,Eidelman:2004wy}
\begin{eqnarray}
  \Omega_{\rm c}^{\rm (WMAP)} h^2 = 0.113^{+0.008}_{-0.009},
\end{eqnarray}
and it is not automatic to realize such density parameter with the
LSP.  It is often the case that some mechanism is needed to enhance
the annihilation of the LSP to realize the LSP dark matter.  Detailed
study shows that the relic density of the LSP can become consistent
with the WMAP value with some particular annihilation process.  In the
so-called mSUGRA models, the parameter regions where the density
parameter of the LSP becomes consistent with the WMAP value are
classified as ``bulk region,'' ``coannihilation
region,'' ``focus-point region,'' ``rapid-annihilation funnels,''
although the annihilation mechanisms used in those regions work
in more general framework.

Once the supersymmetry is discovered, one of the challenge will be to
determine the thermal relic density of the LSP and to see if the LSP
dark matter is plausible
\cite{Battaglia:2004gk,Allanach:2004xn,Bambade:2004tq,Khotilovich:2005gb,
Moroi:2005nc,Carena:2005gc,Birkedal:2005jq,Battaglia:2005ie}.  If
successful, it will give us deeper understandings of our universe up
to the temperature of $O(10\ {\rm GeV})$.  For this purpose, it is
necessary to measure various parameters in the MSSM for the
calculation of the thermal relic density of the LSP.  Such studies
have been also performed for the case of the LHC
\cite{Drees:2000he,Baer:2003wx,Battaglia:2004mp,Polesello:2004qy,
Janot:2004mw,Baer:2004qq}.  Most of those studies have, however,
assumed very simple model of supersymmetry breaking, like the mSUGRA
model, in estimating the expected uncertainties in the reconstructed
value of the thermal relic density of the LSP.  We believe that, once
the superparticles are found, it is necessary to calculate the relic
density in a model-independent way.  Only with the LHC, such a study
seems difficult in many cases.

With the ILC, on the contrary, precise determination of the MSSM
parameters will be possible without assuming any particular model of
SUSY breaking.  Thus, in this case, the ILC will be able to help
reconstructing the relic density of the LSP in many cases.  For
example, in the focus-point case, in \cite{Moroi:2005nc}, it was shown
that the ILC can provide useful and significant information for the
calculation of the thermal relic density of the LSP.

Here, we pay particular attention to another case,
the rapid-annihilation funnels.  In this case,
pair annihilation of the LSP is dominated by the diagram with the
$s$-channel exchange of the CP-odd neutral Higgs boson.  In
particular, in order to enhance the pair annihilation of the lightest
neutralino, $\tan\beta$ parameter becomes large in this case.  If the
LSP dark matter is realized in the rapid-annihilation funnels,
detailed study of the heavy Higgs sector in the large $\tan\beta$ case
is necessary to calculate the annihilation cross section of the LSP.

Strongly motivated by the dark matter physics in the
rapid-annihilation funnels (as well as the study of the Higgs
potential), in this paper, we consider the strategy for studying the
properties of the heavy Higgs bosons at the ILC  assuming an
energy and integrated luminosity of $\sqrt{s}=1$ TeV and $L=1$
ab$^{-1}$.  We concentrate on the case where $\tan\beta$ parameter is
large and see how and how well we can constrain the properties of the
heavy Higgs bosons.  As we will discuss in the following sections,
interactions of the heavy Higgses are well parameterized by five
parameters.  Using pair production processes of the neutral and
charged Higgs bosons, these parameters can be constrained.  If
information from the neutral and charged Higgs production processes is
combined, it will help improving the determination of properties of
the heavy Higgses.   In this paper, we particularly consider how
to combine information from various processes and how well the
resultant constraint can be, using systematic consideration of the
production and decay processes.  Then, we also study how the
results can be used to reconstruct the thermal relic density of the
LSP.

The organization of this paper is as follows.  In Section
\ref{sec:higgs}, we briefly review the properties of the heavy Higgs
bosons.  In Section \ref{sec:ilc}, we discuss how and how well we can
study the properties of the heavy Higgses at the ILC.  In particular,
we estimate the expected errors in the observed values of masses and
widths of the heavy Higgses.  Then, in Section \ref{sec:cdm}, we
discuss the implication of the ILC study of the heavy Higgses to the
reconstruction of the dark matter density.  Section \ref{sec:summary}
is devoted to the conclusions and discussion.

\section{Higgs Sector}
\setcounter{equation}{0}
\label{sec:higgs}

We start by discussing the Higgs bosons in the MSSM and by summarizing
the relevant parameters for our analysis.  As we mentioned, the MSSM
contains two Higgs bosons $H_u$ $({\bf 1},{\bf 2},\frac{1}{2})$ and
$H_d$ $({\bf 1},{\bf 2}^*,-\frac{1}{2})$, where we have shown their
quantum numbers of $SU(3)_C\times SU(2)_L\times U(1)_Y$ gauge group in
the parenthesis.  Both of these Higgs bosons acquire vacuum
expectation values, which are parameterized by
\begin{eqnarray}
\langle H_u^0 \rangle = v \sin\beta,~~~
\langle H_d^0 \rangle = v \cos\beta,
\end{eqnarray}
where $v\simeq 174\ {\rm GeV}$, and the angle $\beta$ is a free
parameter.  After the electroweak symmetry breaking, it is usually the
case that one linear combination of $H_u$ and $H_d$ becomes the
standard-model-like Higgs $\Phi_{\rm SM}$ while the other combination
$\Phi$ contains physical heavy Higgs, which are given by\footnote
{Strictly speaking, mixing in the CP-even neutral Higgs sector differs
from those of other Higgses, and conventionally the mixing in the
CP-even neutral Higgses are parameterized by the parameter $\alpha$
\cite{Gunion:1989we}.  When the heavy Higgses are much heavier than
the lightest Higgs (i.e., in the so-called ``decoupling limit''),
however, it is expected that $\sin(\beta-\alpha)\rightarrow 1$ and
Eq.\ (\ref{Higgs_diag}) holds with good accuracy.  In this paper, we
only consider such a case.}
\begin{eqnarray}
  \left( \begin{array}{c}
    \Phi_{\rm SM} \\ \Phi
  \end{array} \right) =
  \left( \begin{array}{cc}
    \cos\beta & \sin\beta \\
    - \sin\beta & \cos\beta
  \end{array} \right)
  \left( \begin{array}{c}
    H_d \\ H_u^*
  \end{array} \right).
  \label{Higgs_diag}
\end{eqnarray}
Using the mass eigenstates $H$, $A$, and $H^\pm$, the
heavy Higgs doublet $\Phi=(\Phi^0,\Phi^-)$ is expressed as
\begin{eqnarray}
  \Phi^0 = \frac{1}{\sqrt{2}} (H + iA),~~~
  \Phi^\pm = H^\pm.
\end{eqnarray}
In general two Higgs models, masses of $H$, $A$, and $H^\pm$ are
unconstrained.  In the MSSM, however, their masses, $m_H$, $m_A$, and
$m_{H^\pm}$, are quite degenerate within a few GeV or so.  For the
neutral Higgs sector, this fact makes the study difficult.  In our
following study, we do not adopt the theoretical relation among these
masses but we consider how well we can determine these three masses
(in particular, $m_H$ and $m_A$) at the ILC.

\begin{figure}[t]
    \centerline{\epsfxsize=0.5\textwidth\epsfbox{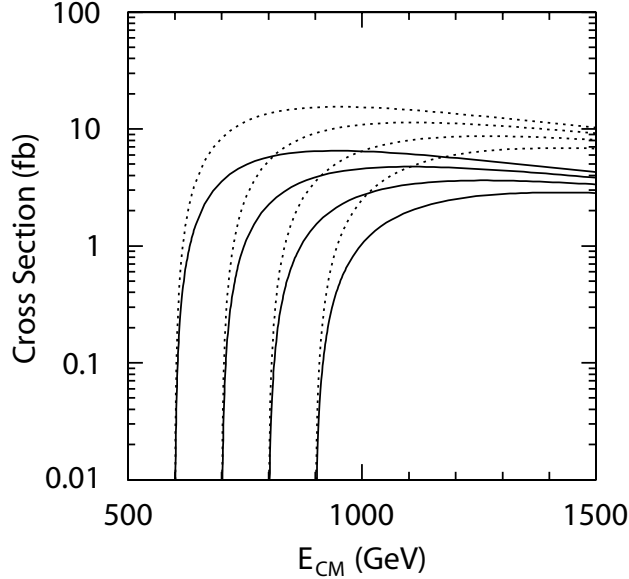}}
    \caption{Total cross sections for the heavy Higgs production
    processes as functions of the center of mass energy $E_{\rm
    CM}=\sqrt{s}$.  The solid lines are for the process
    $e^+e^-\rightarrow AH$ with $m_A=m_H=300\ {\rm GeV}$, $350\ {\rm
    GeV}$, $400\ {\rm GeV}$, and $450\ {\rm GeV}$ from above, while
    the dotted lines are for $e^+e^-\rightarrow H^+H^-$ with
    $m_{H^\pm}= 300\ {\rm GeV}$, $350\ {\rm GeV}$, $400\ {\rm GeV}$,
    and $450\ {\rm GeV}$ from above.}
    \label{fig:cs}
\end{figure}

The productions of the heavy Higgses at the ILC are dominated by the
gauge-boson exchange diagrams.  In Fig.\ \ref{fig:cs}, we show the
cross section for the processes $e^+e^-\rightarrow AH$ and
$e^+e^-\rightarrow H^+H^-$ for several cases.  (We will give the
formula for the cross section in the following section; see Eqs.\ 
(\ref{cs_AH}) and (\ref{cs_HcHc}).)  Here and hereafter, in our
numerical study, we assume un-polarized electron beam.

Decay processes of the heavy Higgses have more model-dependence.  In
the rapid-annihilation funnels, heavy Higgses mostly decay into the
third generation quarks and leptons.  We write the relevant
interaction as\footnote
{Notice that the Yukawa coupling constants $y_b$, $y_t$, and $y_\tau$
are different from those in the superpotential.}
\begin{eqnarray}
  {\cal L}_{\rm Yukawa} =
  y_b \Phi Q_3 b_R^c + y_t \Phi^* i\sigma^2 Q_3 t_R^c 
  + y_\tau \Phi L_3 \tau_R^c + {\rm h.c.},
\end{eqnarray}
where $Q_3$ $({\bf 3},{\bf 2},\frac{1}{6})$, $b_R^c$ $({\bf 3}^*,{\bf
1},\frac{1}{3})$, $t_R^c$ $({\bf 3}^*,{\bf 1},-\frac{2}{3})$, $L_3$
$({\bf 1},{\bf 2},-\frac{1}{2})$, and $\tau_R^c$ $({\bf 1},{\bf 1},1)$
are third-generation quarks and leptons.  In the parameter region we
are interested in, the (fundamental) Yukawa coupling constants in the
superpotential are fairly large.  So, we expect that the radiative
corrections do not significantly affect the relations between those
with Yukawa coupling constants of the heavy Higgses.  If we try to
extract the information about these Yukawa coupling constants from the
observed fermion masses, however, radiative correction may become
important.  At the tree level, the Yukawa coupling constants
are related to the masses of the bottom quark, top quark and $\tau$ as
\begin{eqnarray}
  y_b^{\rm (tree)} = \frac{m_b \tan\beta}{v},~~~
  y_t^{\rm (tree)} = \frac{m_t \cot\beta}{v},~~~
  y_\tau^{\rm (tree)} = \frac{m_\tau \tan\beta}{v}.
  \label{L_Y}
\end{eqnarray}
Thus, once the ``bottom-quark mass'' $m_b$ is well determined by the
study of, for example, the lightest Higgs boson $h$, then the Yukawa
coupling constant for $\tau$ can be also well determined at the tree
level.  It has been pointed out, however, that $y_b$ receives sizable
radiative correction from supersymmetric loop diagrams in particular
at the large $\tan\beta$ region \cite{Hall:1993gn,Carena:1994bv}.
Thus, in determining the interaction of the heavy Higgses, it is
dangerous to adopt the tree level relations.

In our analysis, we assume that several qualitative features of the
large $\tan\beta$ region hold; in particular, we consider the case
where the heavy Higgses decay into third generation quarks and leptons
because of large $y_b$ and $y_\tau$.  We do not, however, rely on the
tree level relations among these Yukawa coupling constants but regard
all the Yukawa couplings (in particular, $y_b$ and $y_\tau$) to be
free parameters which should be experimentally determined.

In the large $\tan\beta$ region, $y_t$ is extremely suppressed and,
hereafter, we neglect its effects.  Then, the neutral Higgses $A$ and
$H$ decay mostly into the $b\bar{b}$ or $\tau\bar{\tau}$ final
states.\footnote
{$H$ may also decay into the gauge-boson or Higgs boson pair, but
those processes are extremely suppressed in the large $\tan\beta$
region that we are interested in.}
The decay rates of the heavy Higgses are given by
\begin{eqnarray}
  \Gamma_{A\rightarrow b\bar{b}} &=& 
  \frac{N_{\rm c}}{16\pi} y_b^2 m_A 
  \left(  1 + \Delta^{\rm (QCD)}_{bb} \right),
  \label{A->bb} \\  
  \Gamma_{A\rightarrow \tau\bar{\tau}} &=&  
  \frac{1}{16\pi} y_\tau^2 m_A,\\
  \Gamma_{H\rightarrow b\bar{b}} &=&  
  \frac{N_{\rm c}}{16\pi} y_b^2 m_H
  \left(  1 + \Delta^{\rm (QCD)}_{bb} \right),\\
  \Gamma_{H\rightarrow \tau\bar{\tau}} &=&  
  \frac{1}{16\pi} y_\tau^2 m_H,\\
  \Gamma_{H^\pm \rightarrow bt} &=&  
  \frac{N_{\rm c}}{16\pi} y_b^2 m_{H^\pm}
  \left( 1 - \frac{m_t^2}{m_{H^\pm}^2} \right)^2
  \left(  1 + \Delta^{\rm (QCD)}_{bt} \right),\\
  \Gamma_{H^\pm \rightarrow \tau\nu_\tau} &=&  
  \frac{1}{16\pi} y_\tau^2 m_{H^\pm},
  \label{Hc->taunu}
\end{eqnarray}
where $N_{\rm c}=3$ is the color factor and we neglect the masses of
the fermions other than the top quark.  Here, we denote the QCD
corrections to the individual decay processes as $\Delta^{\rm
(QCD)}_{bb}$ and $\Delta^{\rm (QCD)}_{bt}$.  Importantly, the QCD
corrections are calculable once the Higgs masses are determined
\cite{Braaten:1980yq,Drees:1990dq}.  Thus, if we can measure the decay
widths and branching ratios of the heavy Higgses, we can constrain the
Yukawa coupling constants.  In addition, the total decay width of the
heavy Higgses are given by
\begin{eqnarray}
  \Gamma_A &=& 
  \Gamma_{A\rightarrow b\bar{b}} 
  + \Gamma_{A\rightarrow \tau\bar{\tau}},
  \\
  \Gamma_H &=&
  \Gamma_{H\rightarrow b\bar{b}} 
  + \Gamma_{H\rightarrow \tau\bar{\tau}},
  \\
  \Gamma_{H^\pm} &=&
  \Gamma_{H^\pm \rightarrow bt}
  + \Gamma_{H^\pm \rightarrow \tau\nu_\tau}.
\end{eqnarray}
As we will see below, in the rapid-annihilation funnels, decay widths
of heavy Higgses are relatively large.  Thus, from the invariant-mass
distribution of the decay products, the decay widths may be measured
at the ILC.

In summary, for the study of the heavy Higgs sector in the large
$\tan\beta$ region, there are five relevant parameters so far: three
masses $m_A$, $m_H$, $m_{H^\pm}$, and two Yukawa coupling constants
$y_b$ and $y_\tau$.  As we will see, the relic density of the LSP can
be also well calculated in the rapid-annihilation funnels once these
parameters are fixed.  In the following, we will discuss how well
these parameters are experimentally determined.

Although our method can be applied to various cases, the whole
parameter space is too large to be studied.  Thus, we choose several
parameter points and see how well we can experimentally measure
parameters at the ILC in those cases.  For simplicity, we generate the
mass spectrum of the MSSM particles by adopting the mSUGRA boundary
condition which is parameterized by the following parameters:
universal scalar mass $m_0$, unified gaugino mass $m_{1/2}$,
tri-linear coupling coefficient $A_0$, $\tan\beta$, and the sign of
the SUSY invariant Higgs mass ${\rm sign}(\mu_H)$.  Although we adopt
the mSUGRA model as the underlying theory to fix the mass spectrum and
mixing parameters, we consider a procedure to experimentally measure
the parameters without assuming the mSUGRA.  Since we are interested
in the implication of such analysis to the study of the dark-matter
physics, we adopt underlying parameters in the rapid-annihilation
funnels.  In Table \ref{table:params}, we list the underlying
parameters which we use.  (We also show the density parameter of the
LSP $\Omega_{\rm LSP}$ with those underlying parameters.)

\begin{table}[t]
  \begin{center}
    \begin{tabular}{lcc}
      \hline\hline
          {} & {Point 1} & {Point 2} \\
          \hline
          $m_0$             & 465.0 GeV & 418.0 GeV   \\
          $m_{1/2}$         & 362.0 GeV & 279.0 GeV   \\
          $A_0$             & 60.0 GeV & $-12.0\ {\rm GeV}$   \\
          $\tan\beta$       & 50     & 48       \\
          $\mu_H$           & 444.5 GeV & 357.7 GeV   \\
          $m_A$             & 400.0 GeV & 350.0 GeV   \\
          $m_H$             & 401.4 GeV & 351.0 GeV   \\
          $m_{H^\pm}$       & 412.9 GeV & 363.6 GeV   \\
          $\Gamma_A$        & 20.7 GeV & 17.1 GeV   \\
          $\Gamma_{H^\pm}$  & 19.7 GeV & 14.9 GeV   \\
          $B_{A\rightarrow b\bar{b}}$ & 0.896 & 0.885  \\
          $m_{\chi^0_1}$    & 144.4 GeV & 109.6 GeV   \\
          $m_{\chi^0_2}$    & 277.3 GeV & 208.3 GeV   \\
          $m_{\chi^0_3}$    & 450.4 GeV & 364.8 GeV   \\
          $m_{\chi^0_4}$    & 467.0 GeV & 383.0 GeV   \\
          $m_{\chi^\pm_1}$  & 277.8 GeV & 208.7 GeV   \\
          $m_{\chi^\pm_2}$  & 467.2 GeV & 383.3 GeV   \\
          $\Omega_{\rm LSP}h^2$ & 0.113 & 0.113  \\
          \hline\hline
    \end{tabular}
    \caption{Underlying parameters to be used in our study.}
    \label{table:params}
  \end{center}
\end{table}

\section{Heavy Higgses at the ILC}
\setcounter{equation}{0}
\label{sec:ilc}

\subsection{Outline}

Now we discuss the study of the properties of the heavy Higgs bosons
at the ILC.  In particular, we are primarily interested in how and how
well we can constrain the five parameters listed in the previous
section by combining the information from the neutral and charged
Higgs production processes.  For this purpose, we adopt several
reasonable approximations to simplify our analysis as we explain
below, instead of studying the details of individual production and
decay processes.

As we mentioned in the previous section, decay rates of the heavy
Higgses are sensitive to the Yukawa coupling constants.  Thus our
strategy is to study branching ratios and the energy distribution of
the final-state fermions produced by the decay of the heavy Higgses.
Since the (total) decay rates are much smaller than the masses of the
heavy Higgses, we can also measure the masses of the heavy Higgses
from the energy distributions of the decay products.

Since we will try to determine the decay rates from the energy
distribution of the final-state particles, we do not use the narrow
width approximation for the heavy Higgs production cross section.  In
our analysis, the following type of the processes play a significant
role: $e^+e^-\rightarrow\varphi_1^*\varphi_2^*\rightarrow
f_1\bar{f}_1'f_2\bar{f}_2'$ where, in our case, $(\varphi_1,
\varphi_2)$ is $(A, H)$ or $(H^+, H^-)$ and $f$'s are quarks and
leptons.  Let us denote the invariant-mass squared of the
$f_i\bar{f}_i'$ system as $s_{f_i\bar{f}_i'}$.  Then, for the
processes we are interested in, the differential cross sections are
well approximated by the following formula\footnote
{For the process we consider (i.e., $e^+e^-\rightarrow
f_1\bar{f}_1'f_2\bar{f}_2'$), there are in fact several diagrams where
some of the fermions are not emitted from the virtual Higgs bosons
(like $e^+e^-\rightarrow f_1^*\bar{f}_1'$, followed by
$f_1^*\rightarrow f_1+\varphi^*$ and $\varphi^*\rightarrow
f_2\bar{f}_2'$).  In the parameter region where
$s_{f_1\bar{f}_1'}^{1/2}$ and $s_{f_2\bar{f}_2'}^{1/2}$ are both close
to the Higgs mass(es), however, we have checked that the cross section
is dominated by the process
$e^+e^-\rightarrow\varphi_1^*\varphi_2^*\rightarrow
f_1\bar{f}_1'f_2\bar{f}_2'$.  Thus, we use Eq.\ (\ref{dsigma/dsds})
for simplicity.}
\begin{eqnarray}
    \frac{d\sigma (s)}{d s_{f_1\bar{f}_1'} d s_{f_2\bar{f}_2'}}
    &=& 
    \frac{\hat{\sigma} (s; s_{f_1\bar{f}_1'}, s_{f_2\bar{f}_2'})}{\pi^2}
    \frac{s_{f_1\bar{f}_1'}^{1/2} 
    \hat{\Gamma}_{\varphi_1\rightarrow f_1\bar{f}_1'}
    (s_{f_1\bar{f}_1'}^{1/2})}{(s_{f_1\bar{f}_1'} - m_{\varphi_1}^2)^2 
    + \Gamma_{\varphi_1}^2 m_{\varphi_1}^2}
    \frac{s_{f_2\bar{f}_2'}^{1/2}
    \hat{\Gamma}_{\varphi_2\rightarrow f_2\bar{f}_2'}
    (s_{f_2\bar{f}_2'}^{1/2})}{(s_{f_2\bar{f}_2'} - m_{\varphi_2}^2)^2 
    + \Gamma_{\varphi_2}^2 m_{\varphi_2}^2}
    \nonumber \\ &&
    + (\varphi_1 \leftrightarrow \varphi_2),
    \label{dsigma/dsds}
\end{eqnarray}
where the second term should be omitted if $f_1$ and $f_2$ are
identical.  Here, $\hat{\sigma}
(s;s_{f_1\bar{f}_1'},s_{f_2\bar{f}_2'})$ is the cross section for the
process $e^+e^-\rightarrow\varphi_1^*\varphi_2^*$ with the masses of
$\varphi_1^*$ and $\varphi_2^*$ being set equal to
$s_{f_1\bar{f}_1'}^{1/2}$ and $s_{f_2\bar{f}_2'}^{1/2}$, respectively.
In addition, $\hat{\Gamma}_{\varphi_i\rightarrow
f_i\bar{f}_i'}(s_{f_i\bar{f}_i'}^{1/2})$ denotes the decay rate of
$\varphi^*_i$ whose mass is identified as $s_{f_i\bar{f}_i'}^{1/2}$;
$\hat{\Gamma}_{\varphi_i\rightarrow
f_i\bar{f}_i'}(s_{f_i\bar{f}_i'}^{1/2})$ can be obtained from Eqs.\ 
(\ref{A->bb}) $-$ (\ref{Hc->taunu}) by replacing the physical mass of
the Higgs by $s_{f_i\bar{f}_i'}^{1/2}$.  Explicit formulae of
$\hat{\sigma}$ will be given in the following subsections.  Notice
that, when $\Gamma_{\varphi_i}\ll m_{\varphi_i}$, the total cross
section becomes
\begin{eqnarray}
  \int d s_{f_1\bar{f}_1'} d s_{f_2\bar{f}_2'}
  \frac{d\sigma}{d s_{f_1\bar{f}_1'} d s_{f_2\bar{f}_2'}}
  \simeq
  ( B_{\varphi_1\rightarrow f_1\bar{f}_1'}
  B_{\varphi_2\rightarrow f_2\bar{f}_2'}
  + B_{\varphi_2\rightarrow f_1\bar{f}_1'}
  B_{\varphi_1\rightarrow f_2\bar{f}_2'} )
  \sigma_{e^+e^-\rightarrow\varphi_1\varphi_2},
\end{eqnarray}
where 
\begin{eqnarray}
  \sigma_{e^+e^-\rightarrow\varphi_1\varphi_2} (s) 
  =
  \hat{\sigma} (s; m_{\varphi_1}^2, m_{\varphi_2}^2).
\end{eqnarray}

As suggested by Eq.\ (\ref{dsigma/dsds}), the distributions of the
invariant masses of the final-state fermions have peaks around
$s_{f_i\bar{f}_i'}^{1/2}\sim m_{\varphi_i}$.  In addition, the
distributions become broader as the decay rate $\Gamma_{\varphi_i}$
becomes larger.  Thus, in the heavy Higgs production processes, the
invariant-mass distributions of the final state fermions have
important information about the masses and the decay widths of the
heavy Higgses.

It should be, however noted that the observed invariant-mass
distributions are deformed by several effects.  One reason is the
energy loss by the neutrino emission.  With the leptonic decays, some
fraction of the initial energy of the bottom quark may be carried away
by energetic neutrinos and the energy of the $b$-jet becomes
underestimated.  Another important effect is from the resolution of
the calorimeters (in particular, the hadronic calorimeter).  In order
to simulate these effects, we perform a Monte-Carlo (MC) analysis.
Here, we first generate parton-level events for a given value of the
collider energy $\sqrt{s}$, which we take to be $1\ {\rm TeV}$
throughout this paper unless otherwise mentioned.  In each event, in
general, momenta of bottom quarks $p^{(0)}_{b_i}$, other lighter
quarks $p^{(0)}_{q_i}$, and leptons $p^{(0)}_{l_i}$ are generated.  If
the final-state partons contain the top quark, its decay (as well as
the subsequent decay of $W^\pm$) is also taken into account at this
state.  Here and hereafter, the superscript $(0)$ is for momenta of
the primary partons before the hadronization and cascade decay.  In
order to consider the energy loss of the bottom quark in the leptonic
decay events, we used ISAJET package \cite{Paige:2003mg} to follow the
hadronization and decay chain of the bottom quark.  For individual
primary bottom quark with (initial) momentum $p_{b_i}^{(0)}$, we
calculate the fraction of the visible energy $f_{b_i}$ after its
hadronization and decay.  Then, the $b$-jet (after the hadronization
and the decay) is treated as the jet with the momentum
$f_{b_i}p_{b_i}^{(0)}$.  In addition, because of the detector
resolution, distribution of the observed energy of the jet is smeared.
In order to determine the observed momentum of the individual jets, we
adopt the energy resolution of the hadronic calorimeter to be
$\sigma_{\rm E}/E=40\ \%/\sqrt{E}$ (with $E$ being the energy of the
jet) \cite{JLC1,Yamashita}.  As a result, with the generated momenta
of the quarks $p^{(0)}_{b_i}$ and $p^{(0)}_{q_i}$, we determine the
observed energy $p_{b_i}$ and $p_{q_i}$.  (We also denote
corresponding observed energy as $E_{b_i}$ and $E_{q_i}$.)

\subsection{Neutral Higgs production}

Let us first consider the neutral Higgs production $e^+e^-\rightarrow
AH$.  In the large $\tan\beta$ region, $A$ and $H$ both dominantly
decay into $b\bar{b}$ and $\tau\bar{\tau}$.  Thus, there are three
types of final states:\footnote
{For simplicity, sometimes we do not distinguish particle and
anti-particle when there is no confusion.}
\begin{eqnarray}
  e^+e^- \rightarrow A^* H^* \rightarrow
  \left\{ \begin{array}{l}
    bbbb \\
    bb\tau\tau \\
    \tau\tau\tau\tau
    \end{array} \right. .
\end{eqnarray}
The cross section for the process $e^+e^-_{L,R}\rightarrow A^* H^*$ is
given by
\begin{eqnarray}
  \hat{\sigma}_{e^+e^-_{L,R}\rightarrow A^* H^*} 
  (s, m_{A^*}^2, m_{H^*}^2) = 
  \frac{s v_{AH}^3}{192\pi}
  \frac{f_{L,R}^2 g_z^2}{(s-m_Z^2)^2} \sin^2(\beta-\alpha),
  \label{cs_AH}
\end{eqnarray}
where $g_Z\equiv\sqrt{g_1^2+g_2^2}$ with $g_1$ and $g_2$ being gauge
coupling constants for the $U(1)_Y$ and $SU(2)_L$ gauge groups,
respectively, $m_Z$ is the $Z$-boson mass, $\alpha$ is the mixing angle
in the CP-even Higgs sector,
\begin{eqnarray}
  v_{AH}^2 = \frac{1}{s}
  \left[ (s-m_{A^*}^2-m_{H^*}^2)^2 
    - 4 m_{A^*}^2 m_{H^*}^2 \right],
  \label{p_3(AH)}
\end{eqnarray}
and, for the left- and right-polarized electron beam,
\begin{eqnarray}
  f_{L} = \frac{g_2^2-g_1^2}{g_z},~~~
  f_{R} = - \frac{g_1^2}{g_z}.
\end{eqnarray}
We take $\sin^2(\beta-\alpha)=1$ since we are interested in the limit
where the heavy Higgses are much heavier than the lightest Higgs.

In our analysis, we use the first two types of events: $bbbb$ ($4b$)
final state and $bb\tau\tau$ final state.  We do not consider
$\tau\tau\tau\tau$ final state because the branching ratios of the
heavy neutral Higgses into the $\tau\tau$ final state are fairly small
($\sim 0.1$) so the number of $4\tau$ events is suppressed.

The branching ratios $B_{A\rightarrow b\bar{b}}$ and $B_{H\rightarrow
b\bar{b}}$ are $\sim 0.9$ so sizable number of $4b$ event is expected.
At the ILC, such an event will be identified by the following
features:
\begin{itemize}
\item 4 $b$-tagged jets.
\item Small missing energy.
\item No isolated leptons.
\end{itemize}

Once observed, the distribution of the invariant mass of two $b$-jets
provides important information about the properties of the heavy Higgs
bosons.  Energy and invariant-mass distributions of the primary
partons are relatively easily calculated.  As we mentioned in the
previous subsection, however, observed distributions are different
from the primary ones because of the leptonic decay of the bottom
quark and also because of the resolution of the calorimeters.  We are
particularly interested in the distribution of the invariant mass of
two $b$-jets which originate from (virtual) neutral Higgs bosons.  In
order to pair the $b$-jets, we use the fact that the masses of the two
(heavy) neutral Higgses are expected to be quite degenerate.
Consequently, the total energy of $b\bar{b}$ system from the virtual
$A$ becomes very close to that from the virtual $H$.  Thus, for the
event $e^+e^-\rightarrow b_1b'_1b_2b'_2$ (where the $b$-jets are
ordered so that the observed energy be
$E_{b_1}<E_{b_2}<E_{b'_2}<E_{b'_1}$), we define the following two
invariant masses:
\begin{eqnarray}
m_{bb,1}^2 = (p_{b_1}+p_{b_1'})^2,~~~
m_{bb,2}^2 = (p_{b_2}+p_{b_2'})^2.
\end{eqnarray}
In general, we can perform statistical analysis based on the
distributions of these two invariant masses.  For simplicity, however,
we use ``averaged'' distribution
\begin{eqnarray}
  \frac{dN_{4b}}{d m_{bb}} \equiv
  \frac{1}{2}  
  \int d m_{bb,1} \left[ \frac{dN_{4b}}{d m_{bb,1} d m_{bb,2}}
    \right]_{m_{bb,2}=m_{bb}}
  + \frac{1}{2}  
  \int d m_{bb,2} \left[ \frac{dN_{4b}}{d m_{bb,1} d m_{bb,2}}
    \right]_{m_{bb,1}=m_{bb}},
  \label{dN(4b)/dm}
\end{eqnarray}
where $N_{4b}$ is the number of $4b$ events.  As we will see, even
from this averaged distribution, we obtain significant constraints on
the masses and decay rates of the neutral Higgses.

In order to determine the distribution given in Eq.\ 
(\ref{dN(4b)/dm}), we calculate the ``transfer functions'' for the
$4b$ events $T_{4b}(m_{bb};s_{b_1\bar{b}_1},s_{b_2\bar{b}_2})$;
$T_{4b}$ is the distribution of the observable $m_{bb}$ for the event
with $s_{b_1\bar{b}_1}$ and $s_{b_2\bar{b}_2}$ being fixed.  In
calculating the transfer function, several kinematical constraints are
taken into account to eliminate standard-model backgrounds.  In the
$4b$ events, all the beam energy is carried away by the bottom quarks.
In addition, for the process $e^+e^-\rightarrow A^*H^*$ followed by
$A^*\rightarrow b_1\bar{b}_1$ and $H^*\rightarrow b_2\bar{b}_2$,
$E_{b_1}+E_{\bar{b}_1}$ becomes close to $E_{b_2}+E_{\bar{b}_2}$ in
most of the cases.  The kinematical cuts based on these can be used to
eliminate some of the backgrounds; here, for the event
$e^+e^-\rightarrow b_1b'_1b_2b'_2$ with
$E_{b_1}<E_{b_2}<E_{b'_2}<E_{b'_1}$, we adopt
\begin{itemize}
\item $960\ {\rm GeV} \leq E_{b_1} + E_{b'_1} + E_{b_2} + E_{b'_2}
    \leq 1040\ {\rm GeV}$.
\item $470\ {\rm GeV} \leq E_{b_1} + E_{b'_1} \leq 530\ {\rm GeV}$ and
    $470\ {\rm GeV} \leq E_{b_2} + E_{b'_2} \leq 530\ {\rm GeV}$.
\item There is no leptonic activity (with energy greater than $25\ 
    {\rm GeV}$), in particular, in the $b$-jets.
\item Invariant masses of any two of the jets are larger than $130\ 
    {\rm GeV}$.
\end{itemize}

With the excellent $b$-tagging capability of the ILC, we expect that
significant fraction of the $b$-jets can be identified.  Tagging of
all four bottom quarks will help eliminating standard-model
backgrounds.  Using the transfer function, we obtain
\begin{eqnarray}
    \frac{dN_{4b}}{dm_{bb}} =
    \epsilon_b^4 L 
    \int ds_{b_1\bar{b}_1} ds_{b_2\bar{b}_2}
    T_{4b} (m_{bb}; s_{b_1\bar{b}_1}, s_{b_2\bar{b}_2})
    \frac
    {d\sigma_{e^+e^-\rightarrow A^*H^*\rightarrow b\bar{b}b\bar{b}}}
    {ds_{b_1\bar{b}_1} ds_{b_2\bar{b}_2}},
\end{eqnarray}
with $L$ being the luminosity.  Here, $\epsilon_b$ is the tagging
efficiency of single bottom quark; we approximate that the $b$-tagging
efficiency is independent of the number of $b$-jets and take
$\epsilon_b=0.7$ in our numerical calculations
\cite{JLC1,NLC,Aguilar-Saavedra:2001rg}.  Then, in the
statistical analysis, we calculate the number of $4b$ events
integrated over some intervals of the invariant mass:
\begin{eqnarray}
  N_{4b}^{(i)} = \int_{m_{bb}^{(i)}}^{m_{bb}^{(i+1)}}
  d m_{bb} \frac{dN_{4b}}{dm_{bb}},
  \label{N_4b^i}
\end{eqnarray}
where $m_{bb}^{(i)}$ and $m_{bb}^{(i+1)}$ are lower and upper bounds
of the $i$-th bin.

Although we imposed several kinematical cuts, there may still remain
backgrounds.  Since we required that the total energy of the
final-state jets is close to $\sqrt{s}$ as well as that 4 $b$-tagged
jets exist, we suppose that the dominant background is from the events
of the type $e^+e^-\rightarrow b\bar{b}b\bar{b}$.  We use COMPHEP
package \cite{Boos:2004kh} to generate such events and, using the
kinematical cuts we discussed before, we estimate the number of
backgrounds.  We found that, however, the number of background from
the process $e^+e^-\rightarrow b\bar{b}b\bar{b}$ is negligibly small
($\sim 0.01\ {\rm event}/5\ {\rm GeV}/1\ {\rm ab}^{-1}$).  There may also
exist another class of backgrounds which arise from the
mis-identification of the $b$-jet.  Study of such background requires
detailed study of the detector effects and we leave such a study for a
future work.\footnote
{Invariant-mass distribution of the $4b$ event has been also studied
in \cite{Battaglia:2004gk}, where it is also shown that the background
for the $4b$ event is well below the signal.}

\begin{figure}[t]
    \centerline{\epsfxsize=0.75\textwidth\epsfbox{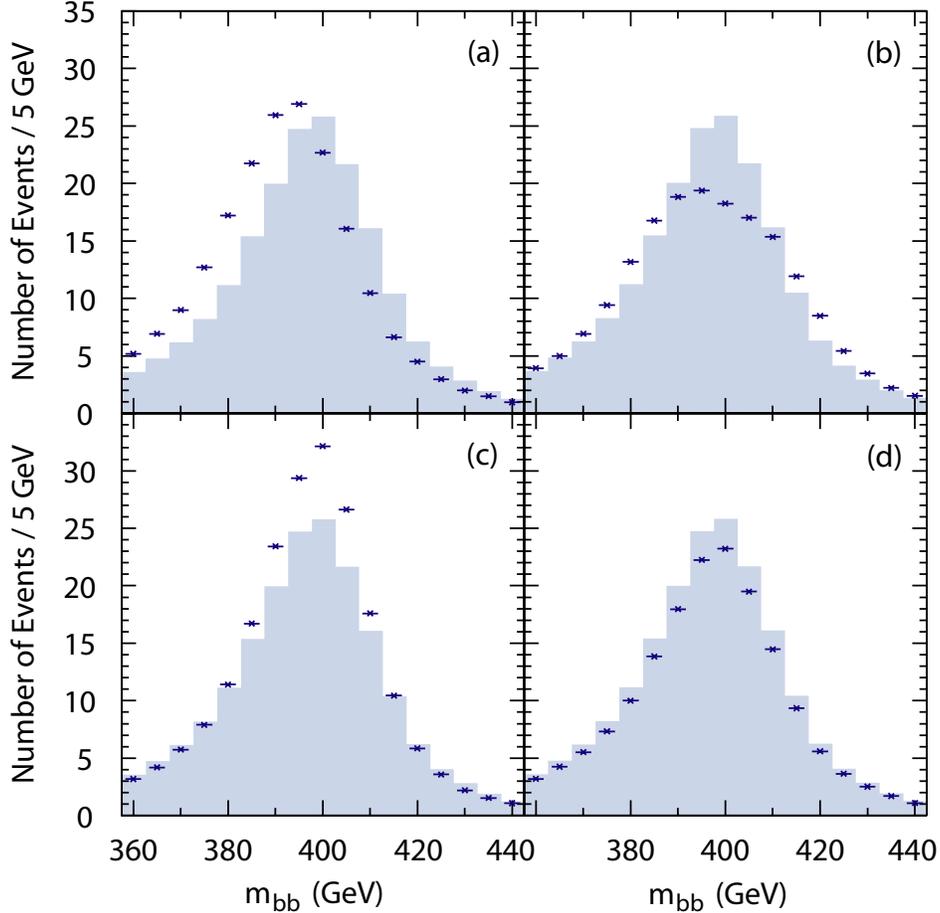}}
    \caption{Number of $4b$ events in each bin (with the width of $5\
    {\rm GeV}$) with $L=1\ {\rm ab}^{-1}$.  The histogram shows the result
    for the Point 1: $\bar{m}_{AH}=400.7\ {\rm GeV}$, $\Delta
    m_{AH}=-0.7\ {\rm GeV}$, $\Gamma_A=20.7\ {\rm GeV}$, and
    $B_{A\rightarrow b\bar{b}}=0.896$.  The short horizontal lines
    with ``$\times$'' are those with one of the four parameters being
    changed: (a) $\bar{m}_{AH}=395\ {\rm GeV}$, (b) $\Delta m_{AH}=10\ 
    {\rm GeV}$, (c) $\Gamma_H=15\ {\rm GeV}$, and (d) $B_{A\rightarrow
    b\bar{b}}=0.85$.}
    \label{fig:N(4b)}
\end{figure}

In Fig.\ \ref{fig:N(4b)}, we show the number of events and backgrounds
in each bin.  Since $A$ and $H$ are quite degenerate, it is convenient
to define the ``averaged mass'' and the ``mass difference'' as
follows:
\begin{eqnarray}
  \bar{m}_{AH} \equiv \frac{1}{2} ( m_A + m_H ),~~~
  \Delta m_{AH} \equiv \frac{1}{2} ( m_A - m_H ) .
\end{eqnarray}
(We will see that error in $\bar{m}_{AH}$ is much smaller than that in
$\Delta m_{AH}$.)  As one can see, the distribution of $m_{bb}$ is
peaked at around $m_{bb}\sim \bar{m}_{AH}$.  In addition, width and
height of the distribution depend on the mass and decay parameters.
Thus, from the observed distribution of $m_{bb}$, mass and decay width
of the neutral Higgses will be determined.

Some fraction of the neutral Higgses also decay into $\tau\bar{\tau}$
pair, so we can also use the events with $b\bar{b}\tau\bar{\tau}$
final state.  Here we only use the hadronic decay mode of $\tau$ to
identify the $\tau$ jets; when the $\tau$ lepton decays hadronically,
we obtain a jet with low multiplicity.  Using this fact, we assume
that the hadrons from $\tau$ can be well identified and distinguished
from the jets from the direct production of quarks.  Then, the
$bb\tau\tau$ event is specified by the following features:
\begin{itemize}
\item 2 $b$-tagged jets ($b$ and $b'$) with the total energy of the
    system being close to $\frac{1}{2}\sqrt{s}$.
\item 2 energetic isolated jets with low multiplicity.
\end{itemize}

Since, in the signal event, two $b$-jets originate from single
(virtual) neutral Higgs, we use the distribution of the invariant mass
of two $b$-jets to determine the masses and widths of the
neutral Higgses:
\begin{eqnarray}
  m_{bb}^2 = (p_{b} + p_{b'})^2.
\end{eqnarray}
For the
$bb\tau\tau$ event, we also calculate the transfer function
$T_{bb\tau\tau}$ with MC simulation.  Then, we obtain
\begin{eqnarray}
    \frac{dN_{bb\tau\tau}}{dm_{bb}} =
    \epsilon_b^2 L B_{\tau \rightarrow {\rm had}}^2
    \int ds_{b\bar{b}} ds_{\tau\bar{\tau}}
    T_{bb\tau\tau}
    (m_{bb}; s_{b_1\bar{b}_1}, s_{b_2\bar{b}_2})
    \frac
    {d\sigma_{e^+e^-\rightarrow A^*H^*\rightarrow 
    b\bar{b}\tau\bar{\tau}}}
    {ds_{b\bar{b}} ds_{\tau\bar{\tau}}},
\end{eqnarray}
where $B_{\tau \rightarrow {\rm had}}\simeq 0.65$
\cite{Eidelman:2004wy} is the hadronic branching ratio of $\tau$.  In
calculating $T_{bb\tau\tau}$, we take account of the following
kinematical cuts:
\begin{itemize}
\item $470\ {\rm GeV} \leq E_{b} + E_{b'} \leq 530\ {\rm GeV}$.
\item In the $b$-jets, there is no leptonic activity with energy
    higher than $25\ {\rm GeV}$.
\item Angle between two jets from $\tau$ leptons is larger than
    $\frac{1}{2}\pi$.
\end{itemize}

For the $bb\tau\tau$ event, the $t\bar{t}$ production process provides
irreducible background; $e^+e^-\rightarrow t\bar{t}$ followed by the
$\tau$-leptonic decay of both the $W^\pm$-bosons produced by the top
decays.  For this process, we generate the events and estimate the
number of background of this type.  We found that the number of events
from this type of background is $(0.1-1)\ {\rm event}/5\ {\rm GeV}/1\ 
{\rm ab}^{-1}$, which is below the number of signal event.  As well as from
the $t\bar{t}$ production process, we may also have backgrounds from
the four-body production process $e^+e^-\rightarrow
b\bar{b}\tau\bar{\tau}$, which is also possible in the standard model.
To see how large it is, we generate such event with COMPHEP package
and estimated the number of background.  We found that the number of
such background events is very small after imposing the kinematical
cuts, $\sim 0.01$ events per each bin (with $L=1\ {\rm ab}^{-1}$) which is
well below the number of signal event.

\begin{figure}[t]
    \centerline{\epsfxsize=0.75\textwidth\epsfbox{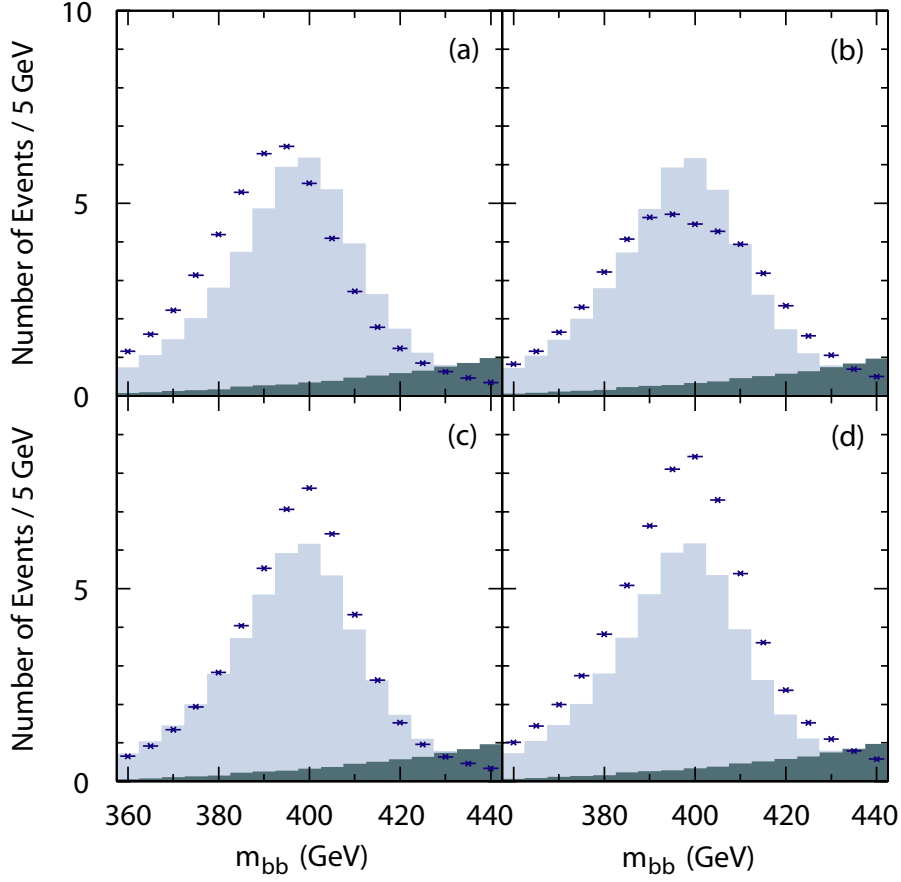}}
    \caption{Same as Fig.\ \ref{fig:N(4b)}, but for the
    $bb\tau\tau$ events.  The background from the $t\bar{t}$
    production process is also shown in the darkly shaded histogram.}
    \label{fig:N(bbtautau)}
\end{figure}

Invariant-mass distribution of the $bb\tau\tau$ event is shown in
Fig.\ \ref{fig:N(bbtautau)}.  As in the case of $4b$ event, the
distribution is peaked at around the heavy neutral Higgs masses.  The
number of $bb\tau\tau$ event is, however, proportional to
$B_{A\rightarrow b\bar{b}} (1-B_{A\rightarrow b\bar{b}})$ and hence is
suppressed compared to that of the $4b$ event.  As a result, $4b$
event is statistically more important than $bb\tau\tau$ event.

Using the above results, we can estimate the expected errors in the
measured values of the physical quantities.  For this purpose, we
first determine the underlying values of the fundamental parameters.
For the neutral Higgs production processes, there are four free
parameters; here we use $m_A$, $m_H$ (or $\bar{m}_{AH}$, $\Delta
m_{AH}$), $\Gamma_A$ and $B_{H\rightarrow b\bar{b}}$ as parameters to
specify the point in the parameter space.  We use the parameter sets
given in Table \ref{table:params} as underlying parameters unless
otherwise stated.  With the underlying parameters, we calculate the
expected number of $4b$ and $bb\tau\tau$ events in each bin, which we
denote $\bar{N}_{4b}^{(i)}$ and $\bar{N}_{bb\tau\tau}^{(i)}$.  Then,
in order to see how well the underlying values can be determined, we
calculate the number of events in each bin, $N_{4b}^{(i)}$ and
$N_{bb\tau\tau}^{(i)}$, for postulated values of $m_A$, $m_H$,
$\Gamma_A$ and $B_{H\rightarrow b\bar{b}}$, and see if
$\bar{N}_{4b}^{(i)}$ and $\bar{N}_{bb\tau\tau}^{(i)}$ are
statistically consistent with $N_{4b}^{(i)}$ and
$N_{bb\tau\tau}^{(i)}$.  Here, we define the $\delta\chi^2$ variable
for the neutral Higgs production as
\begin{eqnarray}
  \delta \chi^2_{\rm N} = 
  \sum_i 
  \frac{(\bar{N}_{4b}^{(i)}-N_{4b}^{(i)})^2}
       {{N}_{4b}^{(i)} + N_{4b,{\rm BG}}^{(i)}}
  + \sum_i
  \frac{(\bar{N}_{bb\tau\tau}^{(i)}-N_{bb\tau\tau}^{(i)})^2}
  {N_{bb\tau\tau}^{(i)} + N_{bb\tau\tau,{\rm BG}}^{(i)}},
  \label{chi2_N}
\end{eqnarray}
where $N_{4b,{\rm BG}}^{(i)}$ and $N_{bb\tau\tau,{\rm BG}}^{(i)}$ are
numbers of backgrounds for the $4b$ and $bb\tau\tau$ events,
respectively.

\begin{figure}[t]
    \centerline{\epsfxsize=0.75\textwidth\epsfbox{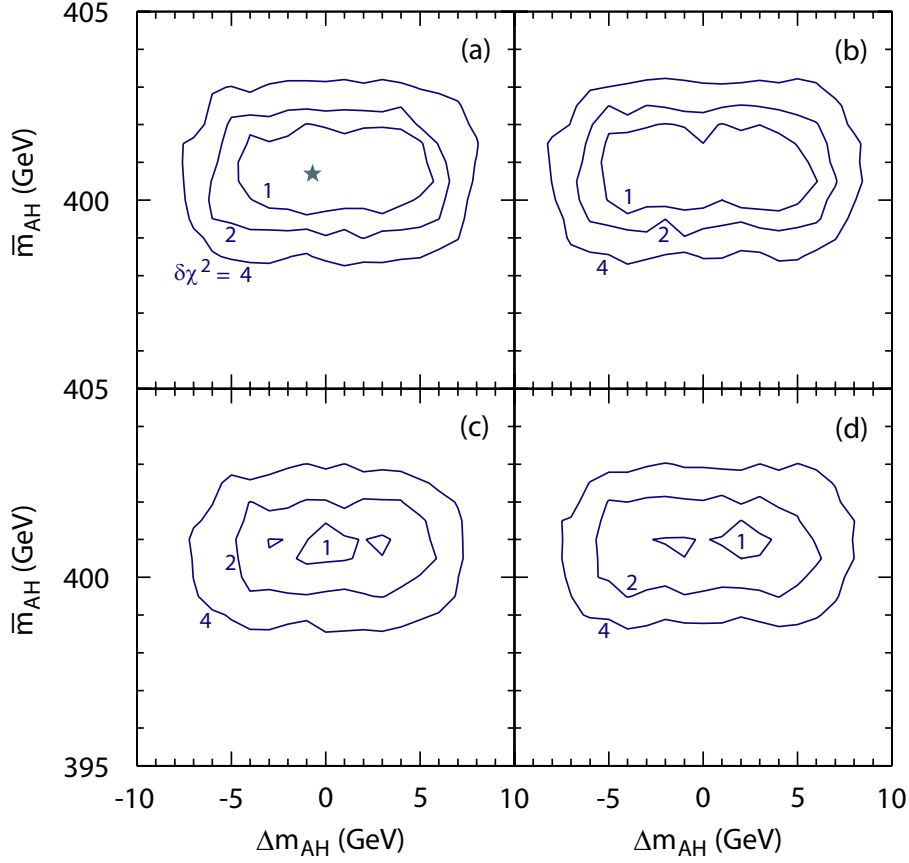}}
    \caption{Contours of constant $\delta\chi^2_{\rm N}$ on $\Delta
    m_{AH}$ vs.\ $\bar{m}_{AH}$ planes; $\delta\chi^2_{\rm N}=1$, $2$,
    and $4$ (with $L=1\ {\rm ab}^{-1}$) from inside and the ``star'' on the
    figure indicates the underlying point.  As the underlying point,
    we take the Point 1 in Table \ref{table:params}.  For the
    calculation of $N_{4b}^{(i)}$ and $N_{bt\tau\nu_\tau}^{(i)}$ in
    Eq.\ (\ref{chi2_N}), we take $\Gamma_A$ and $B_{A\rightarrow
    b\bar{b}}$ to be (a) $\Gamma_A=20.7\ {\rm GeV}$ and
    $B_{A\rightarrow b\bar{b}}=0.896$ (underlying values), (b)
    $\Gamma_A=19.7\ {\rm GeV}$ and $B_{A\rightarrow b\bar{b}}=0.896$,
    (c) $\Gamma_A=20.7\ {\rm GeV}$ and $B_{A\rightarrow
    b\bar{b}}=0.88$, (d) $\Gamma_A=19.7\ {\rm GeV}$ and
    $B_{A\rightarrow b\bar{b}}=0.88$.}
    \label{fig:dm_mbar}
\end{figure}

With $\delta\chi^2_{\rm N}$, we estimate how well we can determine the
physical quantities.  Since $\delta\chi^2_{\rm N}$ depends on four
independent parameters, we fix two of them and show the behaviors of
$\delta\chi^2_{\rm N}$ on several two-dimensional hyperplanes.  First,
we consider the constraints on the masses of the neutral Higgses.  For
this purpose, we consider the hyperplane with $\Gamma_A$ and
$B_{A\rightarrow b\bar{b}}$ being fixed and show the contours of
constant $\delta\chi^2_{\rm N}$ on $\Delta m_{AH}$ vs.\ $\bar{m}_{AH}$
plane. In Fig.\ \ref{fig:dm_mbar}, we show the results.  In
particular, in Fig.\ \ref{fig:dm_mbar}(a), we show the behavior of
$\delta\chi^2_{\rm N}$ on the hyperplane with $\Gamma_A$ and
$B_{A\rightarrow b\bar{b}}$ being equal to their underlying values.
As one can see, if we consider the contour of $\delta\chi^2_{\rm
N}=1$, error of $\bar{m}_{AH}$ is $\sim 1\ {\rm GeV}$ while that of
$\Delta m_{AH}$ is $\sim 6\ {\rm GeV}$.  We also show the contours of
constant $\delta\chi^2_{\rm N}$ on other plane.  We can see that the
contours for $\delta\chi^2_{\rm N}=2$ and $4$ are almost unchanged
while the region with $\delta\chi^2_{\rm N}<1$ is reduced.
From the fact that the error in $\bar{m}_{AH}$ is significantly
smaller than that in $\Delta m_{AH}$, we expect that the error in the
determination of the masses of heavy neutral Higgses is dominantly
from that of $\Delta m_{AH}$.

\begin{figure}[t]
    \centerline{\epsfxsize=0.75\textwidth\epsfbox{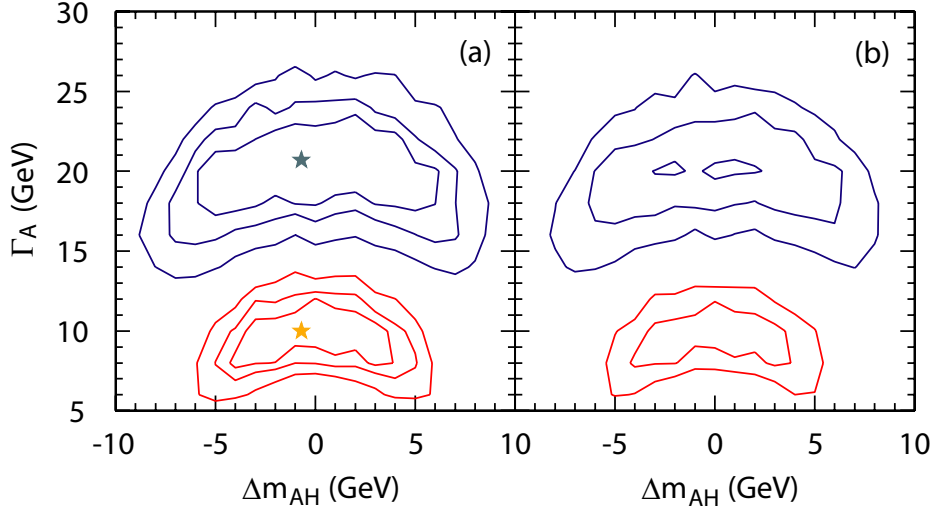}}
    \caption{Contours of constant $\delta\chi^2_{\rm N}$ on $\Delta
    m_{AH}$ vs.\ $\Gamma_A$ planes; $\delta\chi^2_{\rm N}=1$, $2$, and
    $4$ (with $L=1\ {\rm ab}^{-1}$) from inside.  We fix $\bar{m}_{AH}$ and
    $B_{A\rightarrow b\bar{b}}$ to be (a) $\bar{m}_{AH}=400.7\ {\rm
    GeV}$ and $B_{A\rightarrow b\bar{b}}=0.896$ (the underlying
    values), and (b) $\bar{m}_{AH}=400.7\ {\rm GeV}$ and
    $B_{A\rightarrow b\bar{b}}=0.88$.  For the upper contours, the
    underlying values are given by the Point 1. For the lower
    contours, the underlying value of $\Gamma_A$ is taken to be $10\ 
    {\rm GeV}$.  (In (b), there is no countour for $\delta\chi^2_{\rm
    N}=1$ for the case of $\Gamma_A=10\ {\rm GeV}$.)}
    \label{fig:dm_gammaA}
\end{figure}

\begin{figure}[t]
    \centerline{\epsfxsize=0.75\textwidth\epsfbox{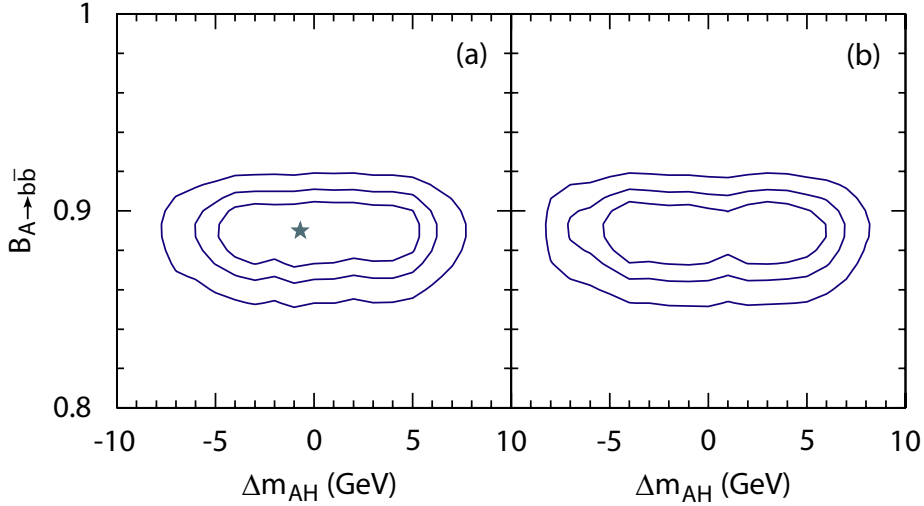}}
    \caption{Contours of constant $\delta\chi^2_{\rm N}$ for 
    the Point 1 on $\Delta m_{AH}$ vs.\ $B_{A\rightarrow b\bar{b}}$
    planes; $\delta\chi^2_{\rm N}=1$, $2$, and $4$ (with $L=1\ {\rm
    ab}^{-1}$) from inside.  Here, $\bar{m}_{AH}$ and $\Gamma_A$ are fixed
    to be (a) $\bar{m}_{AH}=0.7\ {\rm GeV}$ and $\Gamma_A=20.7\ {\rm
    GeV}$ (the underlying values), and (b) $\bar{m}_{AH}=0.7\ {\rm
    GeV}$ and $\Gamma_A=19.7\ {\rm GeV}$.}
    \label{fig:dm_B}
\end{figure}

Next, let us consider the uncertainties in other quantities.  In Fig.\ 
\ref{fig:dm_gammaA}, we show the contours of constant
$\delta\chi^2_{\rm N}$ on the hyperplane of $\bar{m}_{AH}$ and
$B_{A\rightarrow b\bar{b}}$ being fixed.  From this figure, we can see
that the uncertainty in $\Gamma_A$ is $2-3\ {\rm GeV}$ when the
underlying value is $\Gamma_A\simeq 20\ {\rm GeV}$.  In fact, the
uncertainties are sensitive to the underlying value of $\Gamma_A$.  To
see this, we also show the results for the case where the underlying
value of $\Gamma_A$ is $10\ {\rm GeV}$ (with other underlying
parameters being unchanged).  In this case, we can see that the errors
in $\Delta m_{AH}$ and $\Gamma_A$ are reduced.

Using the $\delta\chi^2_{\rm N}$ variable, we can also constrain the
branching ratios.  In Fig.\ \ref{fig:dm_B}, we plot the contours of
constant $\delta\chi^2_{\rm N}$ on $\Delta m_{\rm N}$ vs.\ 
$B_{A\rightarrow b\bar{b}}$ plane.  We can see that $B_{A\rightarrow
b\bar{b}}$ can be constrained at the level of $\sim 5\ \%$ and hence
we can conclude that $B_{A\rightarrow b\bar{b}}$ can be well
determined.

\subsection{Charged Higgs production}

Next, we consider the charged Higgs events.  The useful decay modes in
the large $\tan\beta$ case are $H^\pm\rightarrow bt$ and
$\tau\nu_\tau$.  Thus, in the charged Higgs events, relevant final
states are
\begin{eqnarray}
  e^+e^- \rightarrow H^{+*} H^{-*} \rightarrow
  \left\{ \begin{array}{l}
    btbt \\
    b t \tau \nu_\tau \\
    \tau \nu_\tau \tau \nu_\tau
    \end{array} \right. .
\end{eqnarray}
We expect significant amount of irreducible backgrounds for the
$\tau\nu_\tau\tau\nu_\tau$ event in particular from the $W^+W^-$
production.  Thus, we do not consider this mode.

First, we consider the $bt\tau\nu_\tau$ final state.  After the decay
of the top quark, we obtain $W^\pm$-boson and $b$.  In order to
determine the mass of the charged Higgs, we only use the hadronic
decay of the $W^\pm$-boson.  In addition, for $\tau$, we again use
only the hadronic decay mode to identify the $\tau$-lepton events.
Then, the relevant $bt\tau\nu_\tau$ event has the following features:
\begin{itemize}
\item 2 $b$-tagged jets ($b_1$ and $b_2$).
\item 2 non-$b$-like jets ($q_1$ and $q_2$).
\item 1 energetic isolated jet with low multiplicity.
\end{itemize}

To eliminate backgrounds, we also impose several kinematical
constraints.  Since the event is dominated by the back-to-back
production of the charged Higgs bosons (which are almost on-shell),
total energy of the $bt$ system is close to $\frac{1}{2}\sqrt{s}$ in
most of the signal events.  In addition, if we consider the system
consisting of $q_1$, $q_2$ and one of the $b$-jet, its invariant mass
becomes close to $m_t$.  (We may be able to use the fact that the
invariant mass of the system consisting of $q_1$ and $q_2$ system is
close to $m_W$ although, in our numerical analysis, we do not take
into account such constraint.)

In the $bt\tau\nu_\tau$ event, all the hadronic activities except the
isolated hadrons from $\tau$ are from one of the charged Higgs, so the
invariant mass of such system contain information about the mass and
decay width of $H^\pm$.  Thus, we define
\begin{eqnarray}
  m_{bt}^2 \equiv ( p_{b_1} + p_{b_2} + p_{q_1} + p_{q_2} )^2,
\end{eqnarray}
and calculate the distribution of $m_{bt}$ by MC analysis.  In deriving
the distribution of this invariant mass, we impose the following cuts:
\begin{itemize}
\item $470\ {\rm GeV} \leq E_{b_1}+E_{b_2}+E_{q_1}+E_{q_2}\leq 530\
  {\rm GeV}$.
\item $150\ {\rm GeV} \leq m_{b_1q_1q_2} \leq 200\ {\rm GeV}$ or $150\
  {\rm GeV} \leq m_{b_2q_1q_2} \leq 200\ {\rm GeV}$, where 
  $m_{b_iq_1q_2}^2\equiv (p_{b_i}+p_{q_1}+p_{q_2})^2$. 
\item No leptonic activity (with energy larger than $25\ {\rm GeV}$)
  in the $b$-jets.
\end{itemize}
Then, we calculate the transfer function $T_{bt\tau\nu_\tau}$ to
estimate
\begin{eqnarray}
  \frac{dN_{bt\tau\nu_\tau}}{d m_{bt}} =
  \epsilon_b^2 L B_{\tau\rightarrow {\rm had}} 
  B_{W^\pm \rightarrow q\bar{q}}
  \int ds_{bt} ds_{\tau\nu_\tau}
  T_{bt\tau\nu_\tau} (m_{bt}; s_{bt}, s_{\tau\nu_\tau})
  \frac
      {d\sigma_{e^+e^-\rightarrow H^{+*}H^{-*}\rightarrow bt\tau\nu_\tau}}
      {ds_{bt} ds_{\tau\nu_\tau}} ,
\end{eqnarray}
where $B_{W^\pm \rightarrow q\bar{q}}\simeq 0.676$
\cite{Eidelman:2004wy} is the branching ratio of the hadronic decay of
$W^\pm$-boson.  Here, the cross section for $e^+e^-\rightarrow
H^{+*}H^{-*}$ is given by
\begin{eqnarray}
  \hat{\sigma}_{e^+e^-_{L,R}\rightarrow H^{+*}H^{-*}} 
  (s; m_{H^{+*}}^2,m_{H^{-*}}^2) = 
  \frac{s v_{H^\pm}^3}{48\pi} 
  \left( \frac{e^2}{s} +  f_{L,R} \frac{g_2^2-g_1^2}{g_z}
    \frac{1}{s-m_Z^2} \right)^2,
    \label{cs_HcHc}
\end{eqnarray}
where $e$ is the electric charge, and $v_{H^\pm}$ is obtained from Eq.\
(\ref{p_3(AH)}) by replacing $m_{A^*}\rightarrow m_{H^{+*}}$ and
$m_{H^*}\rightarrow m_{H^{-*}}$.

\begin{figure}[t]
    \centerline{\epsfxsize=0.75\textwidth\epsfbox{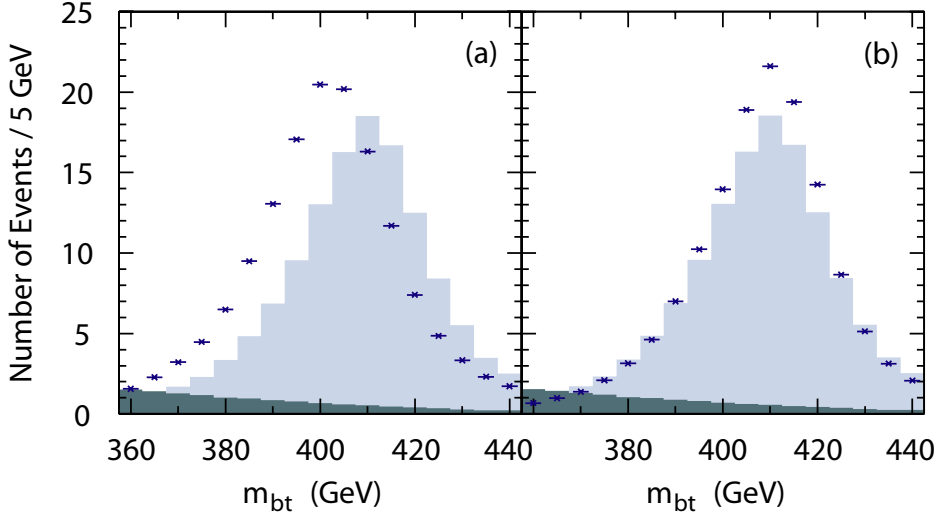}}
    \caption{Number of $bt\tau\nu_\tau$ events in each bin.  
    The shaded histogram shows the result for the Point 1:
    $m_{H^\pm}=412.9\ {\rm GeV}$, $\Gamma_{H^\pm}=19.7\ {\rm GeV}$,
    and $B_{H^\pm\rightarrow bt}=0.885$.  The short horizontal lines
    with ``$\times$'' are those with one of the parameters being
    changed: (a) $m_{H^\pm}=405\ {\rm GeV}$, and (b)
    $\Gamma_{H^\pm}=15\ {\rm GeV}$. The background from the $t\bar{t}$
    production process is also shown in the darkly shaded histogram.}
    \label{fig:histHc}
\end{figure}

In Fig.\ \ref{fig:histHc}, we plot the distribution of the invariant
mass $m_{bt}$ for several cases.  First of all, as one can see, the
distribution function is peaked at around $\sim m_{H^\pm}$.  In
addition, the distribution becomes broader as the decay width
$\Gamma_{H^\pm}$ becomes larger.  Notice that the number of
$bt\tau\nu_\tau$ event is approximately proportional to
$B_{H^\pm\rightarrow bt}(1-B_{H^\pm\rightarrow bt})$. 

For the $bt\tau\nu_\tau$ process, there exists irreducible background
from the $t\bar{t}$ pair production process, if one of the $W^\pm$
boson decays hadronically while the other decays into $\tau$ and
$\nu_\tau$.  We have estimated the number of the background from
$t\bar{t}$ pair production, generating such events.  The result is
also shown in Fig.\ \ref{fig:histHc}; the number of background is well
below the number of signal event.  We also calculated the background
from the standard-model process of $e^+e^-\rightarrow bt\tau\nu_\tau$
with the COMPHEP package with eliminating the contribution from the
process $e^+e^-\rightarrow t\bar{t}$; we found that the number of
background event of this type is extremely small.

As a charged Higgs production event, $b\bar{t}\bar{b}t$ final state
event is also available.  In this case, we expect four $b$-tagged jets
and other hadronic and/or leptonic activities in the final state (from
the decay of the $W^\pm$ bosons).  In the $btbt$ events, however, the
reconstruction of the invariant mass of the $bt$ system of one side
seems challenging because of the combination error or overlap of the
jets (although it may be possible with some careful analysis).  Thus,
in this paper, we only use the total number of the $btbt$ event to
constrain the branching ratio, using the fact that the number of the
$btbt$ event is approximately proportional to $B_{H^\pm\rightarrow
bt}^2$.  Here, we estimate the number of $btbt$ event as
\begin{eqnarray}
    N_{btbt} = \epsilon_b^4 B_{H^\pm\rightarrow bt}^2
    L \sigma_{e^+e^-\rightarrow H^+H^-}.
\end{eqnarray}

As in the case of the neutral Higgs, we calculate the $\delta\chi^2$
variable; for the $bt\tau\nu_\tau$ event, we calculate the number of
events falling into each bin, which is classified by $m_{bt}$.  The
number of event in the $i$-th bin $N_{bt\tau\nu_\tau}^{(i)}$ is
calculated as a function of independent parameters (in the charged
Higgs case, $m_{H^\pm}$, $\Gamma_{H^\pm}$, and $B_{H^\pm\rightarrow
bt}$).  For the $btbt$ event, we use only the total number of events
$N_{btbt}$.  Then, we define
\begin{eqnarray}
  \delta \chi^2_{\rm C} = 
  \sum_i 
  \frac{(\bar{N}^{(i)}_{bt\tau\nu_\tau}-N_{bt\tau\nu_\tau}^{(i)})^2}
  {N_{bt\tau\nu_\tau}^{(i)}+N_{bt\tau\nu_\tau,{\rm BG}}^{(i)}}
  + \frac{(\bar{N}_{btbt}-N_{btbt})^2}{N_{btbt}+N_{btbt,{\rm BG}}},
\end{eqnarray}
where the second terms in the denominators are the numbers of
background events.

\begin{figure}[t]
    \centerline{\epsfxsize=0.75\textwidth\epsfbox{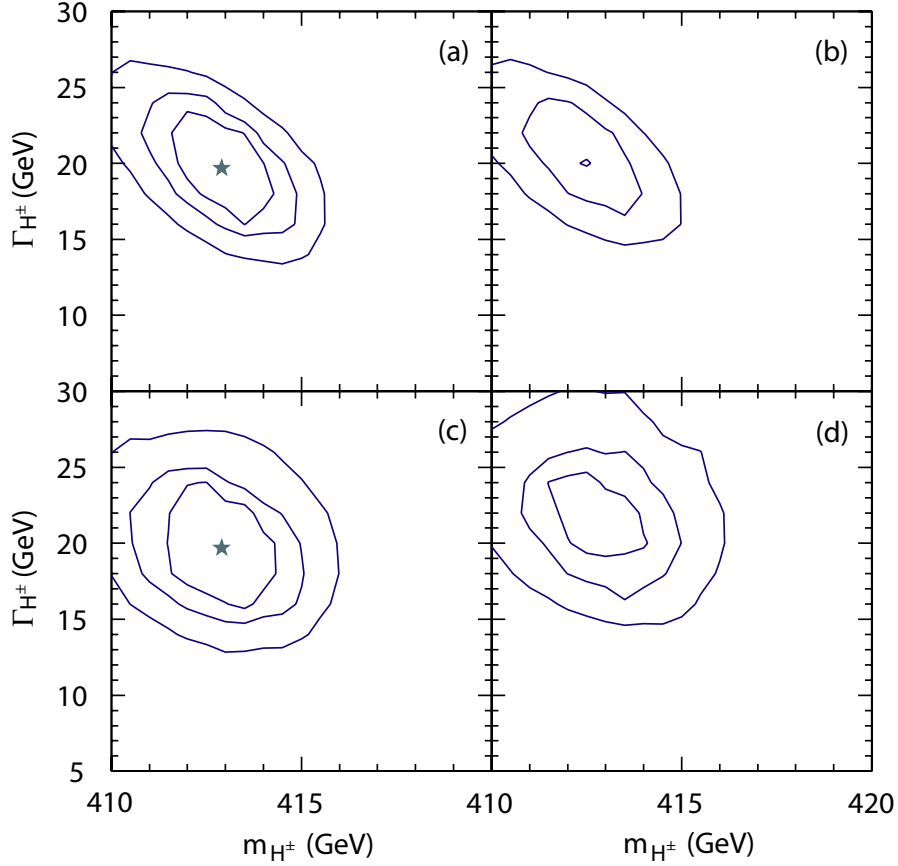}}
    \caption{Contours of constant $\delta\chi^2_{\rm C}$ on
    $m_{H^\pm}$ vs.\ $\Gamma_{H^\pm}$ planes; $\delta\chi^2_{\rm C}=1$,
    $2$, and $4$ (with $L=1\ {\rm ab}^{-1}$) from inside.  The branching
    ratio is fixed to be $B_{H^\pm\rightarrow bt}=0.885$ for (a) and
    (c), and $B_{H^\pm\rightarrow bt}=0.87$ for (b) and (d).  Upper
    two figures ((a) and (b)) are with $tbtb$ mode while the lower
    ones ((c) and (d)) are without $tbtb$ mode.}
    \label{fig:x2c_mcwd}
\end{figure}

\begin{figure}[t]
    \centerline{\epsfxsize=0.75\textwidth\epsfbox{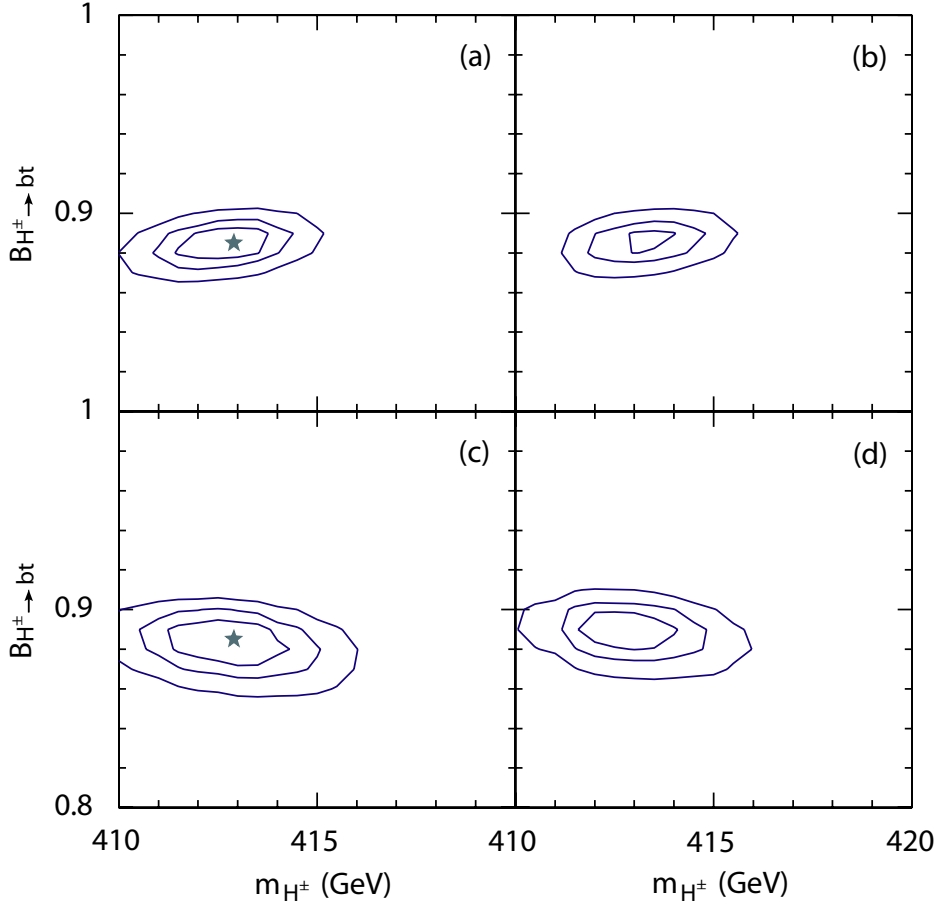}}
    \caption{Contours of constant $\delta\chi^2_{\rm C}$ on
    $m_{H^\pm}$ vs.\ $B_{H^\pm\rightarrow bt}$ planes;
    $\delta\chi^2_{\rm C}=1$, $2$, and $4$ (with $L=1\ {\rm ab}^{-1}$) from
    inside.  Here, we take $\Gamma_{H^\pm}=19.7\ {\rm GeV}$ for (a)
    and (c), and $\Gamma_{H^\pm}=17.0\ {\rm GeV}$ for (b) and (d).
    Upper two figures are with $tbtb$ mode while the lower ones are
    without $tbtb$ mode.}
    \label{fig:x2c_mcbr}
\end{figure}

In Fig.\ \ref{fig:x2c_mcwd}, we show the contours of constant
$\delta\chi^2_{\rm C}$ on $m_{H^\pm}$ vs.\ $\Gamma_{H^\pm}$ planes.
We can see that, from the contour of $\delta\chi^2_{\rm C}=1$, the
uncertainty in $m_{H^\pm}$ is $1-2\ {\rm GeV}$ while that in
$\Gamma_{H^\pm}$ is $\sim 3\ {\rm GeV}$.  We can also see that the
inclusion of the $tbtb$ mode can help reducing the allowed region on
the $m_{H^\pm}$ vs.\ $\Gamma_{H^\pm}$ plane.  In Fig.\ 
\ref{fig:x2c_mcbr}, we also show the contours of constant
$\delta\chi^2_{\rm C}$ on $m_{H^\pm}$ vs.\ $B_{H^\pm\rightarrow bt}$
planes.  We can see that the branching ratio $B_{H^\pm\rightarrow bt}$
is also well constrained by the study of the charged Higgs production
processes.

\subsection{Combined results}

Now, we combine the results obtained in the previous subsections.
For this purpose, we define the total $\delta\chi^2$ variable as
\begin{eqnarray}
  \delta\chi^2_{\rm tot} = \delta\chi^2_{\rm N} 
  + \delta\chi^2_{\rm C}.
\end{eqnarray}
This quantity depends on five parameters (as well as on the underlying
parameters); we use $m_A$, $m_H$, $m_{H^\pm}$, $\Gamma_A$, and
$B_{A\rightarrow b\bar{b}}$ as free parameters.  Once these parameters
are fixed, all the Yukawa coupling constants (as well as the QCD
correction factors) are determined in the large $\tan\beta$ region and
the number of the neutral and charged Higgs production events can be
calculated.  As we saw in the previous subsections, averaged mass of
the neutral Higgses $\bar{m}_{AH}$, the charged Higgs mass $m_{H^\pm}$
and the branching ratios are relatively well determined. Thus, in this
subsection, we concentrate on the uncertainties in the remaining two
parameters which are most important for the calculation of the relic
density of the LSP, i.e., $\Delta m_{AH}$ and $\Gamma_A$.

\begin{figure}
    \centerline{\epsfxsize=0.5\textwidth\epsfbox{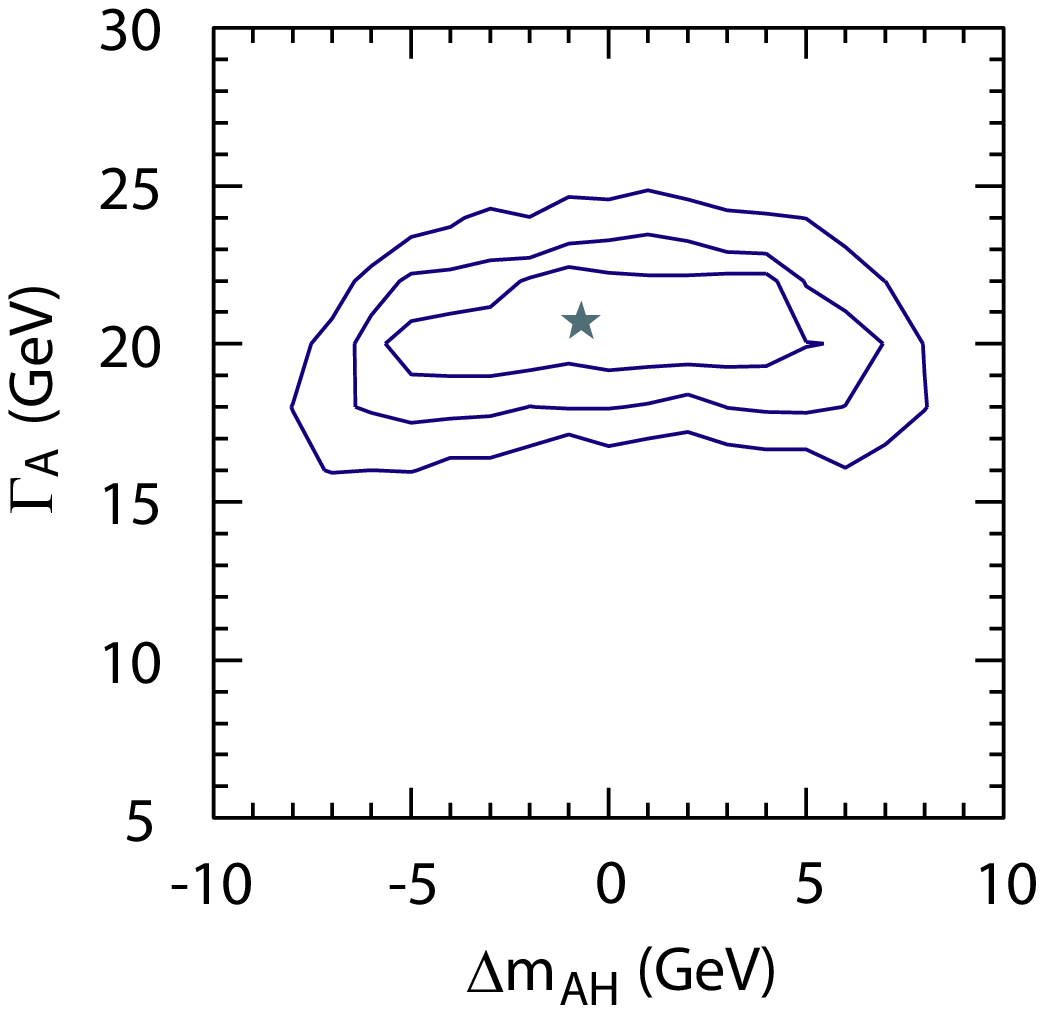}}
    \caption{Contours of constant $\delta\chi^2_{\rm tot}$ for Point 1
    on $\Delta m_{AH}$ vs.\ $\Gamma_A$ plane; $\delta\chi^2_{\rm
    tot}=1$, $2$, and $4$ (with $L=1\ {\rm ab}^{-1}$) from inside.
    $\bar{m}_{AH}$, $m_{H^\pm}$, and $B_{A\rightarrow b\bar{b}}$ are
    fixed to be the underlying values.}
    \label{fig:chi2tot_1}
%
    \vspace{10mm}
    \centerline{\epsfxsize=0.5\textwidth\epsfbox{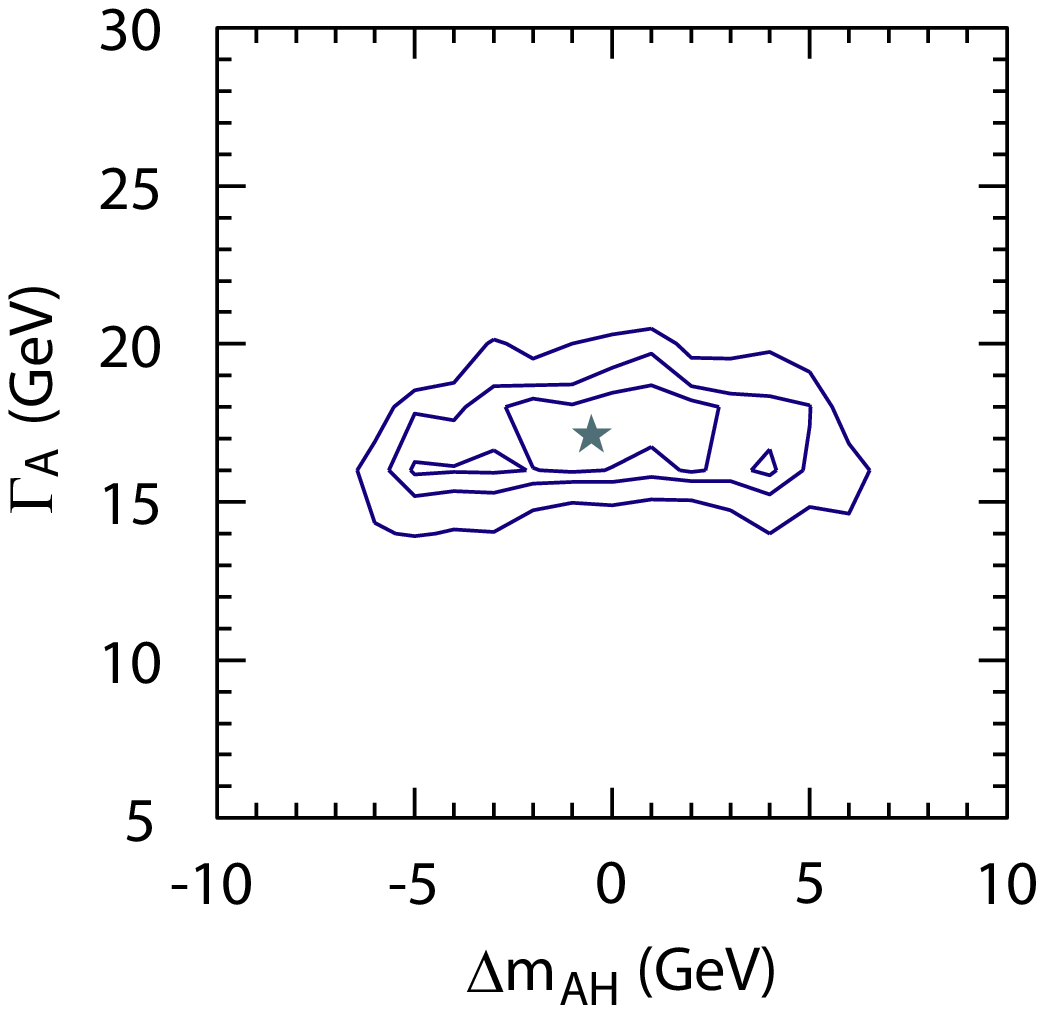}}
    \caption{Same as Fig.\ \ref{fig:chi2tot_1}, but for the
    Point 2.}
    \label{fig:chi2tot_2}
\end{figure}

In Fig.\ \ref{fig:chi2tot_1}, we plot the contours of constant
$\delta\chi^2_{\rm tot}$ for Point 1 on $\Delta m_{AH}$ vs.\ 
$\Gamma_A$ plane, on which $\bar{m}_{AH}$, $m_{H^\pm}$, and
$B_{A\rightarrow b\bar{b}}$ are fixed to be the underlying values.  As
one can see, constraint on the $\Delta m_{AH}$ vs.\ $\Gamma_A$ plane
becomes more stringent as we add the information from the charged
Higgs production events.  Compared with Fig.\ \ref{fig:dm_gammaA}, we
can see that the errors in $\Delta m_{AH}$ and $\Gamma_A$ are reduced
by including the information from the charged Higgs productions.  We
also performed the same study for the Point 2, and the result is given
in Fig.\ \ref{fig:chi2tot_2}.

Before closing this section, let us comment on the determination of
the $\tan\beta$ parameter.  Since the decay widths of the heavy
Higgses are sensitive to $\tan\beta$, we can obtain information about
$\tan\beta$ from the decay widths of the heavy Higgses
\cite{Gunion:2002ip}.  In order to derive the value of $\tan\beta$,
however, we need to relate the Yukawa coupling constants to the
observed masses of the (third-generation) fermions.  For the bottom
quark, this is difficult since the radiative correction from the
supersymmetric diagrams may significantly affect the bottom quark
mass.  Expecting that the radiative correction to the mass of $\tau$
is small enough, however, we can use the relation
\begin{eqnarray}
    \tan\beta = \frac{y_\tau v}{m_\tau}.
\end{eqnarray}
Then, using the decay rate of the CP-odd Higgs boson, for example, we
obtain
\begin{eqnarray}
    \tan\beta =
    \sqrt{\frac{16\pi \Gamma_A (1 - B_{A\rightarrow b\bar{b}})}{m_A}}
    \frac{v}{m_\tau},
\end{eqnarray}
and hence
\begin{eqnarray}
    \frac{\delta \tan\beta}{\tan\beta} \simeq
    \frac{1}{2} \left[ 
        \left( \frac{\delta \Gamma_A}{\Gamma_A} \right)^2
        + \left( \frac{\delta m_A}{m_A} \right)^2
        + \left( \frac{\delta B_{A\rightarrow b\bar{b}}}
            {1 - B_{A\rightarrow b\bar{b}}} \right)^2
    \right]^{1/2}.
\end{eqnarray}
As we have seen in the previous subsections, uncertainties in
$\Gamma_A$ and $B_{A\rightarrow \tau\bar{\tau}}=1-B_{A\rightarrow
b\bar{b}}$ are both $10-20\ \%$ level, while that in $m_A$ is smaller
($\sim 1\ \%$).  Thus, in this case, the $\tan\beta$ parameter can be
determined with the accuracy of $10\ \%$ or so.

\section{Reconstruction of Dark Matter Density}
\setcounter{equation}{0}
\label{sec:cdm}

Now, we are at the position to consider the possibility of
reconstructing the dark matter density in the rapid-annihilation
funnels.  In the parameter region we are interested in, pair
annihilation of the lightest neutralino (which is the LSP) is
dominated by the process with the $s$-channel exchange of the
pseudo-scalar Higgs boson $A$.  The cross section for this process is
calculated as\footnote
{Strictly speaking, in Eq.\ (\ref{sigma*v}), we have to take account
of the QCD correction to the decay rate $\Gamma_A$
when $s_{\chi^0_1\chi^0_1}^{1/2}$ differs from
$m_A$.  Such a QCD correction is calculable and, in this paper, we
neglect such a correction.}
\begin{eqnarray}
    \sigma_{\chi^0_1\chi^0_1\rightarrow {\rm all}} v_{\rm rel} 
    \simeq
    \frac{2 y_{A\chi^0_1\chi^0_1}^2 \Gamma_A}{m_A}
    \frac{s_{\chi^0_1\chi^0_1}}
    {(s_{\chi^0_1\chi^0_1} - m_A^2)^2 + \Gamma_A^2 m_A^2}.
    \label{sigma*v}
\end{eqnarray}
where $s_{\chi^0_1\chi^0_1}$ is the invariant mass of two LSPs in the
initial state, and $y_{A\chi^0_1\chi^0_1}$ is the coupling between 
$A$ and the lightest neutralino, which is given by
\begin{eqnarray}
    y_{A\chi^0_1\chi^0_1} &= &
    \frac{1}{2} \left(
        g_1 \left[ U_{\chi^0} \right]_{11}
        - g_2 \left[ U_{\chi^0} \right]_{12} \right)
    \left(
        \sin\beta \left[ U_{\chi^0} \right]_{13}
        - \cos\beta \left[ U_{\chi^0} \right]_{14} \right).
    \label{y_Achichi}
\end{eqnarray}
Here, $U_{\chi^0}$ is the unitary matrix which diagonalizes the
neutralino mass matrix which is given by
\begin{eqnarray}
  {\cal M}_{0} &=& \left( \begin{array}{cccc}
    -m_{\rm G1} & 0 & - g_1 v \cos\beta & g_1 v \sin\beta \\
    0 & -m_{\rm G2} & g_2 v \cos\beta & - g_2 v \sin\beta \\
    - g_1 v \cos\beta & g_2 v \cos\beta & 0 & \mu_H \\
    g_1 v \sin\beta & - g_2 v \sin\beta & \mu_H & 0
  \end{array} \right),
  \label{m_neutralino}
\end{eqnarray}
with $m_{\rm G1}$, $m_{\rm G2}$, and $\mu_H$ being the gaugino masses
for $U(1)_Y$ and $SU(2)_L$ gauge groups and SUSY invariant Higgs mass,
respectively.  Thus, in the rapid-annihilation funnels, dominant
contribution to the pair annihilation cross section of the lightest
neutralino is calculated once $m_A$ and $\Gamma_A$ as well as
$y_{A\chi^0_1\chi^0_1}$ are determined.  In the previous section, we
have seen that, by the study of the heavy Higgses, $m_A$ and
$\Gamma_A$ are constrained fairly well.

For the determination of $y_{A\chi^0_1\chi^0_1}$, we need the
mixing matrix $U_{\chi^0}$ (as well as $\tan\beta$) which is
calculated from the neutralino mass matrix.  Importantly, the
neutralino mass matrix depends on the gaugino masses and the SUSY
invariant Higgs mass which can be determined once the chargino and
neutralino masses are experimentally measured. At the ILC,
measurements of these masses can be performed if the charginos and
neutralinos are kinematically accessible.  In particular, by studying
the production processes at the threshold region (i.e., by the
threshold scan), some of the masses of the charginos and neutralinos
can be determined with the accuracy of  $\sim 50\ {\rm MeV}$
\cite{Aguilar-Saavedra:2001rg}.  In addition, from the kinematics of
the decay products of the chargino and neutralinos (as well as
sfermions, if kinematically accessible), mass of the LSP (i.e., the
lightest neutralino) is also determined with $\delta m_{\chi^0_1}\sim
50\ {\rm MeV}$ \cite{Aguilar-Saavedra:2001rg}.  Thus, if all the
charginos and neutralinos are seen at the ILC, it will give us enough
information to precisely determine $y_{A\chi^0_1\chi^0_1}$.  If some
of the charginos or the neutralinos are too heavy to be experimentally
produced, it becomes rather difficult to determine the neutralino
mixing parameters.  In such a case, we may have to perform some global
fit using all the masses of the superparticles.  Of course, some
information from the LHC may be also used.  Since our main concern is
to study the properties of the heavy Higgses, we do not consider the
detail of the parameters in the neutralino sector.  Instead, we assume
that, at the ILC, all the charginos and neutralinos can be produced
and that their masses can be precisely measured.   If the masses
of all the charginos and neutralinos are measured with the accuracy of
$\sim 50\ {\rm MeV}$, the uncertainty of $y_{A\chi^0_1\chi^0_1}$ is
expected to be $O(0.01\ \%)$ so that the dominant uncertainty in the
reconstructed value of $\Omega_{\rm LSP}$ is from $m_A$ and
$\Gamma_A$.  

\begin{figure}
    \centerline{\epsfxsize=0.5\textwidth\epsfbox{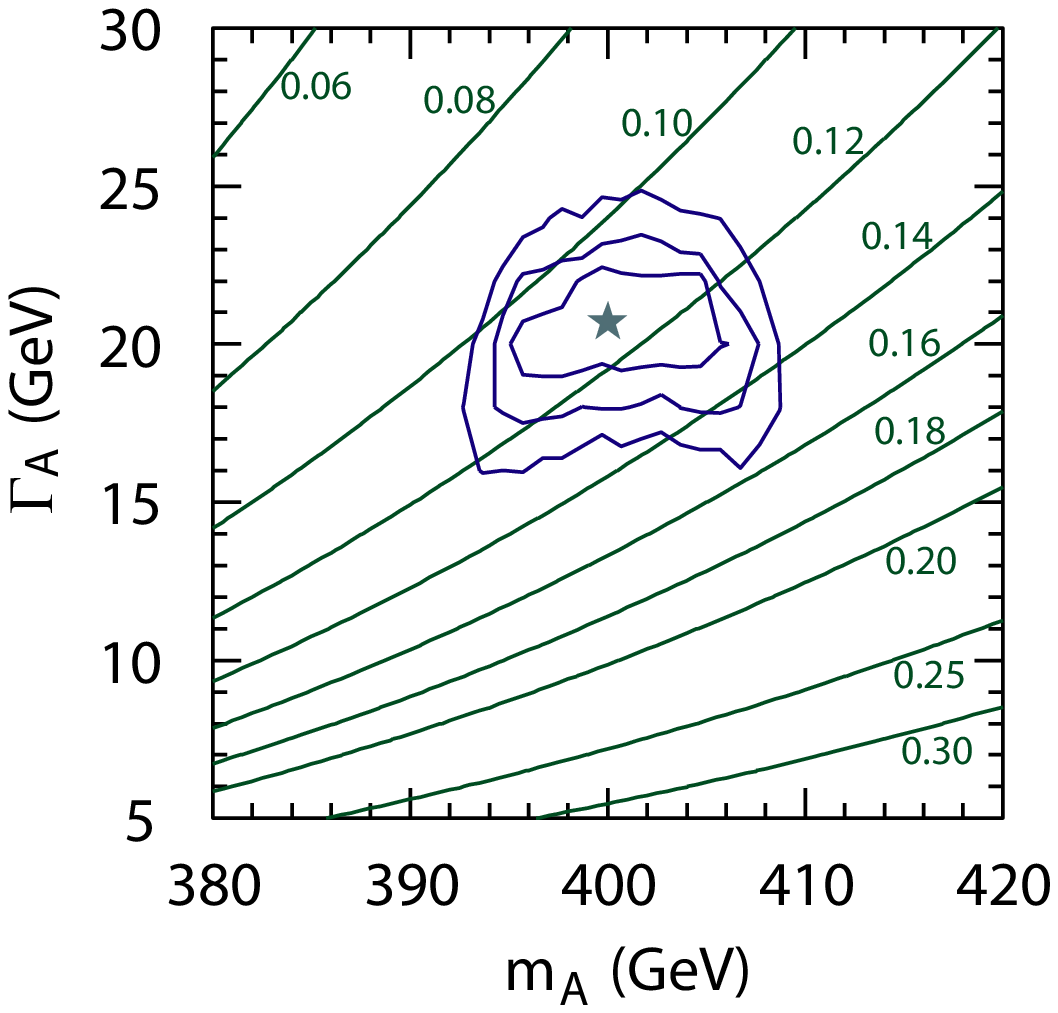}}
    \caption{Contours of constant thermal relic density of the LSP on the 
    $m_A$ vs.\ $\Gamma_A$ plane for the Point 1.  The numbers in the
    figure are the value of $\Omega_{\rm LSP}h^2$.  Contours
    surrounding the star indicate the expected constraints on the
    $m_A$ vs.\ $\Gamma_A$ plane from the ILC ($\delta\chi^2_{\rm
    tot}=1$, $2$, and $4$ from inside, with $L=1\ {\rm ab}^{-1}$).}
    \label{fig:omega1}
%
    \vspace{10mm}
    \centerline{\epsfxsize=0.5\textwidth\epsfbox{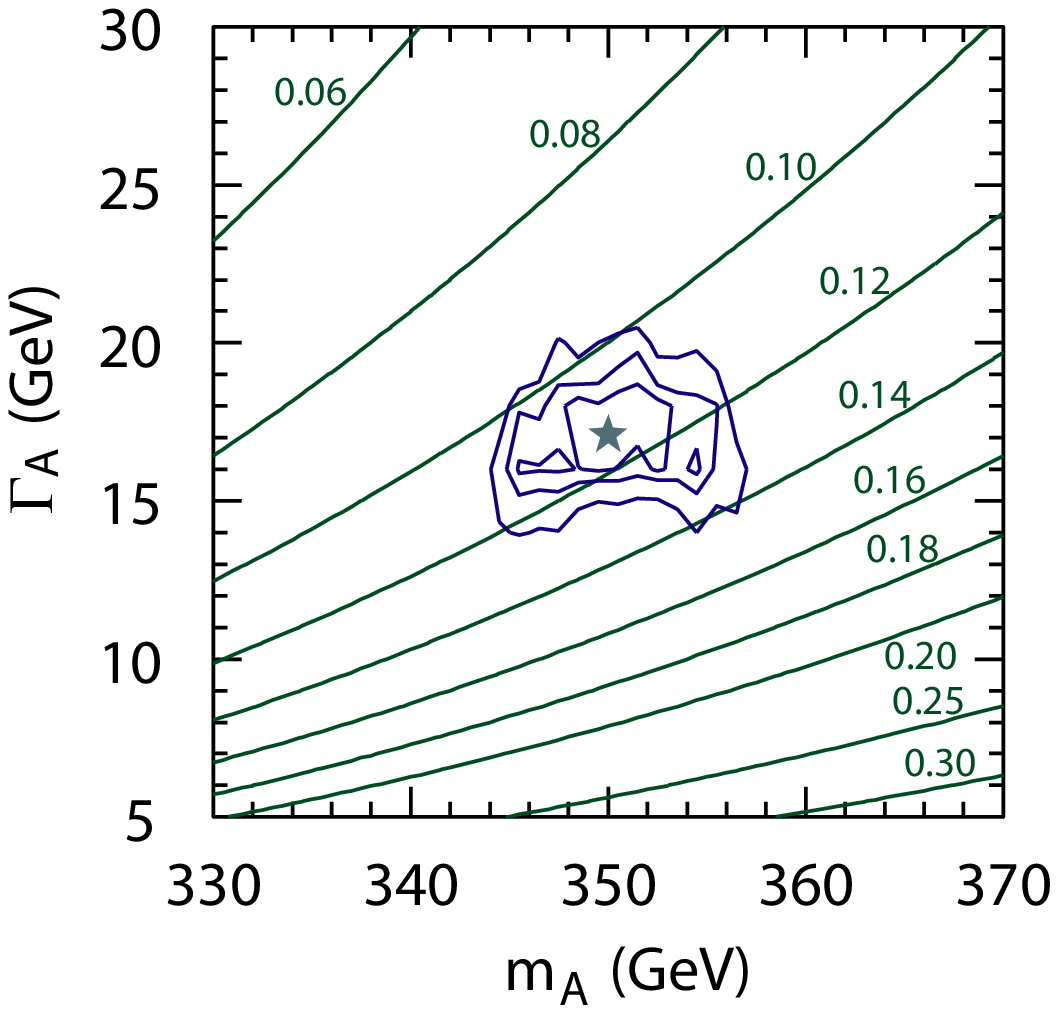}}
    \caption{Same as Fig.\ \ref{fig:omega1} but for Point 2.}
    \label{fig:omega2}
\end{figure}

To see how well we can reconstruct $\Omega_{\rm LSP}$, in Figs.\ 
\ref{fig:omega1} and \ref{fig:omega2}, we plot theoretically
calculated value of $\Omega_{\rm LSP}$ on $m_A$ vs.\ $\Gamma_A$ plane.
(In our numerical calculation, we use the DarkSUSY package
\cite{Gondolo:2004sc} to calculate $\Omega_{\rm LSP}$.)  Here, we
fixed $y_{A\chi^0_1\chi^0_1}$ and $m_{\chi^0_1}$ as their underlying
values and calculated $\Omega_{\rm LSP}$ as a function of $m_A$ and
$\Gamma_A$.  On the same figure, we also show the constraint on the
$m_A$ vs.\ $\Gamma_A$ plane expected from the ILC.  Here, assuming
that the dominant error of $m_A$ is from $\Delta m_{AH}$, we neglected
the uncertainty of $\bar{m}_{AH}$.  From these figures, we can see
that the uncertainty of $\Omega_{\rm LSP}$ is $10 - 20\ \%$, 
which is comparable to the uncertainty of the dark-matter density
determined from the WMAP data.

We have not discussed possible errors of $\Omega_{\rm LSP}$
originating from parameters other than $m_A$, $\Gamma_A$, and
$y_{A\chi^0_1\chi^0_1}$.  In order to calculate the total pair
annihilation cross section of the LSP, we also have to take account of
other processes like the $s$-channel $H^0$ exchange and $t$-channel
sfermion exchange processes.  In the rapid-annihilation funnels,
however, effects of these processes are subdominant. (We have checked
that processes other than the CP-odd Higgs exchange change the value
of $\Omega_{\rm LSP}$ at most a few \% in Points 1 and 2.)  In
addition, more importantly, once the properties of the heavy Higgses
and the sfermions are determined at the ILC, cross sections for these
subdominant processes are also calculated.  Thus, we expect that the
uncertainty in $\Omega_{\rm LSP}$ from the processes other than the
CP-odd Higgs exchange is less than a few \% and that the dominant
errors are from $m_A$ and $\Gamma_A$.

\section{Conclusions and Discussion}
\setcounter{equation}{0}
\label{sec:summary}

We have considered the capability of the $e^+e^-$ linear collider for
studying the properties of the heavy Higgs bosons in the
supersymmetric standard model at the ILC.  We concentrated on the
large $\tan\beta$ region which is motivated, in particular, by
explaining the dark-matter density of the universe (i.e., so-called
``rapid-annihilation funnels'').  We perform a systematic analysis to
estimate expected uncertainties in the determination of the masses and
widths of the heavy Higgs bosons.

With the study of the invariant-mass distributions of the jets in the
final state, we have seen that the masses, widths, and the branching
ratios of the heavy Higgses are well constrained. Compared to the
averaged value $\bar{m}_{AH}=\frac{1}{2}(m_A+m_H)$, mass difference of
the neutral Higgses, $\Delta m_{AH}=\frac{1}{2}(m_A-m_H)$, is more
difficult to measure.  Consequently, if we try to experimentally
determine the masses of the neutral heavy Higgses, uncertainty in
$\Delta m_{AH}$ becomes the significant source of the error.

In this paper, our primary purpose was to point out the strategy for
the systematic study of the heavy Higgs bosons at the ILC and to
estimate the expected uncertainties in the measurements of their
masses and widths.  Thus, we assumed that the $b$-tagging efficiency
and the energy resolution of the detector is well understood and we
did not consider systematic errors from these.  In addition, for the
background, we take account only of the dominant physics background.
When the study suggested in this paper will be performed, these points
should be studied in more detail; in order to realize the detailed and
precise study of the heavy Higgs bosons, the following will be
necessary:
\begin{itemize}
\item Good understanding and high efficiency of the $b$-tagging.
\item High resolution of, in particular, hadron calorimeter.
\item Good understanding of the backgrounds.
\end{itemize}

We have also discussed the implication of the study of the heavy
Higgses to the calculation of the relic density of the LSP.  In the
rapid-annihilation funnels, the dominant pair annihilation process of
the LSP is through the $s$-channel exchange of the CP-odd Higgs boson.
In this case, the mass and width of the CP-odd Higgs boson should be
determined for the precise calculation of the relic density of the
LSP.  We have seen that, if the LSP dark matter is realized in the
rapid-annihilation funnels, the the dark matter density can be
reconstructed with a very good accuracy of $10-20\ \%$ (see Figs.\ 
\ref{fig:omega1} and \ref{fig:omega2}).  If the uncertainty in the
mass or the width of the CP-odd Higgs can be reduced, we will have a
better determination of $\Omega_{\rm LSP}$.  In particular, the
dominant source of the uncertainty in $m_A$ is from the determination
of the mass difference of two neutral heavy Higgses.  If the radiative
correction to the Higgs potential is well understood, then the mass
difference $\Delta m_{AH}$ may be theoretically calculated.  In this
case, uncertainty in $m_A$ becomes smaller and a better determination
of $\Omega_{\rm LSP}$ is expected.  In addition, we have also assumed
that the radiative corrections to the pair annihilation processes of
the LSP will become well studied by the time when the superparticles
as well as the heavy Higgses are produced at the ILC.  Although most
of the radiative corrections are calculable, many of them have not
been calculated yet.  In order for the precise theoretical calculation
of $\Omega_{\rm LSP}$, such a study will be very important.

In conclusion, we have seen that the ILC is very useful not only for
studying the properties of the new particles in particle-physics
models beyond the standard model (in this case, the heavy Higgse in
the sypersymmetric models) but also to have deeper insights into the
evolution of the universe.  In particular, the ILC may help answering
one of the most serious misteries in cosmology, the origin of the dark
matter of the universe.  If the reconstruction of the dark matter
density will be successful, it will provide us a better understanding
of our universe up to the temperature of $O(10\ {\rm GeV})$.

\section*{Acknowledgment}

This work is supported in part by the 21st century COE program,
``Exploring New Science by Bridging Particle-Matter Hierarchy.''  The
work of T.M. is also supported by the Grants-in Aid of the Ministry of
Education, Science, Sports, and Culture of Japan No.\ 15540247.


\begin{thebibliography}{99}

\bibitem{JLC1}
    S.~Matsumoto {\it et al.} [JLC Group],
    ``JLC-1,'' 
    KEK Report 92-16 (1992).

\bibitem{NLC}
    S.~Kuhlman {\it et al.} [The NLC ZDR Design Group and The NLC
    Physics Working Group],
    ``Physics and Technology of the Next Linear Collider,''
    BNL 52-502 (1996).

\bibitem{Aguilar-Saavedra:2001rg}
    J.~A.~Aguilar-Saavedra {\it et al.}  
    [ECFA/DESY LC Physics Working Group],
    arXiv:hep-ph/0106315.

\bibitem{Snowmass}
    For the recent activities, see also the Web page of ``2005
    International Linear Collider Physics and Deterctor Workshop and
    Second ILC Accelerator Worksnop'' (2005, Snowmass, Colorado), {\tt
    http://alcpg2005.colorado.edu}.

\bibitem{Djouadi:1996pj}
  A.~Djouadi, J.~Kalinowski, P.~Ohmann and P.~M.~Zerwas,
  Z.\ Phys.\ C {\bf 74}, 93 (1997).

\bibitem{Gunion:1996cc}
  J.~F.~Gunion and J.~Kelly,
  Phys.\ Rev.\ D {\bf 56}, 1730 (1997).

\bibitem{Feng:1996xv}
  J.~L.~Feng and T.~Moroi,
  Phys.\ Rev.\ D {\bf 56}, 5962 (1997).

\bibitem{Barger:2000fi}
  V.~D.~Barger, T.~Han and J.~Jiang,
  Phys.\ Rev.\ D {\bf 63}, 075002 (2001).

\bibitem{Gunion:2002ip}
  J.~F.~Gunion, T.~Han, J.~Jiang and A.~Sopczak,
  Phys.\ Lett.\ B {\bf 565}, 42 (2003).

\bibitem{Desch:2004yb}
  K.~Desch, T.~Klimkovich, T.~Kuhl and A.~Raspereza,
  arXiv:hep-ph/0406229.


\bibitem{Baer:2002gm}
  H.~Baer  {\it et al.},
  JHEP {\bf 0207}, 050 (2002).

\bibitem{Ellis:2003cw}
  J.~R.~Ellis, K.~A.~Olive, Y.~Santoso and V.~C.~Spanos,
  Phys.\ Lett.\ B {\bf 565}, 176 (2003).

\bibitem{Baer:2003yh}
  H.~Baer and C.~Balazs,
  JCAP {\bf 0305}, 006 (2003).

\bibitem{Chattopadhyay:2003xi}
  U.~Chattopadhyay, A.~Corsetti and P.~Nath,
  Phys.\ Rev.\ D {\bf 68}, 035005 (2003).

\bibitem{Lahanas:2003yz}
  A.~B.~Lahanas and D.~V.~Nanopoulos,
  Phys.\ Lett.\ B {\bf 568}, 55 (2003).

\bibitem{Baer:2003wx}
  H.~Baer {\it et al.},
  JHEP {\bf 0306}, 054 (2003).

\bibitem{Battaglia:2003ab}
  M.~Battaglia {\it et al.},
  Eur.\ Phys.\ J.\ C {\bf 33}, 273 (2004).

\bibitem{Belanger:2004ag}
  G.~Belanger {\it et al.},
  Nucl.\ Phys.\ B {\bf 706}, 411 (2005).

\bibitem{Baltz:2004aw}
  E.~A.~Baltz and P.~Gondolo,
  JHEP {\bf 0410}, 052 (2004).

\bibitem{Bennett:2003bz}
    C.~L.~Bennett {\it et al.},
    Astrophys.\ J.\ Suppl.\  {\bf 148} 1 , (2003).

\bibitem{Eidelman:2004wy}
  S.~Eidelman {\it et al.}  [Particle Data Group],
  Phys.\ Lett.\ B {\bf 592}, 1 (2004).

\bibitem{Battaglia:2004gk}
  M.~Battaglia,
  arXiv:hep-ph/0410123.

\bibitem{Allanach:2004xn}
    B.~C.~Allanach, G.~Belanger, F.~Boudjema and A.~Pukhov,
    JHEP {\bf 0412}, 020 (2004).

\bibitem{Bambade:2004tq}
  P.~Bambade, M.~Berggren, F.~Richard and Z.~Zhang,
  arXiv:hep-ph/0406010.

\bibitem{Khotilovich:2005gb}
  V.~Khotilovich, R.~Arnowitt, B.~Dutta and T.~Kamon,
  Phys.\ Lett.\ B {\bf 618}, 182 (2005).

\bibitem{Moroi:2005nc}
  T.~Moroi, Y.~Shimizu and A.~Yotsuyanagi,
  arXiv:hep-ph/0505252.

\bibitem{Carena:2005gc}
  M.~Carena {\it et al.},
  arXiv:hep-ph/0508152.

\bibitem{Birkedal:2005jq}
  A.~Birkedal {\it et al.},
  arXiv:hep-ph/0507214.

\bibitem{Battaglia:2005ie}
  M.~Battaglia and M.~E.~Peskin,
  arXiv:hep-ph/0509135.

\bibitem{Drees:2000he}
  M.~Drees {\it et al.},
  Phys.\ Rev.\ D {\bf 63}, 035008 (2001).

\bibitem{Battaglia:2004mp}
  M.~Battaglia, I.~Hinchliffe and D.~Tovey,
  J.\ Phys.\ G {\bf 30}, R217 (2004).


\bibitem{Polesello:2004qy}
  G.~Polesello and D.~R.~Tovey,
  JHEP {\bf 0405}, 071 (2004).


\bibitem{Janot:2004mw}
  P.~Janot  [CMS Collaboration],
  arXiv:hep-ph/0406275.

\bibitem{Baer:2004qq}
  H.~Baer, A.~Belyaev, T.~Krupovnickas and J.~O'Farrill,
  JCAP {\bf 0408}, 005 (2004).

\bibitem{Gunion:1989we}
  See, for example,
  J.~F.~Gunion, H.~E.~Haber, G.~L.~Kane and S.~Dawson,
  ``The Higgs Hunter's Guide,'' SCIPP-89/13.

\bibitem{Hall:1993gn}
  L.~J.~Hall, R.~Rattazzi and U.~Sarid,
  Phys.\ Rev.\ D {\bf 50}, 7048 (1994).

\bibitem{Carena:1994bv}
  M.~Carena, M.~Olechowski, S.~Pokorski and C.~E.~M.~Wagner,
  Nucl.\ Phys.\ B {\bf 426}, 269 (1994).

\bibitem{Braaten:1980yq}
  E.~Braaten and J.~P.~Leveille,
  Phys.\ Rev.\ D {\bf 22}, 715 (1980).

\bibitem{Drees:1990dq}
  M.~Drees and K.~i.~Hikasa,
  Phys.\ Lett.\ B {\bf 240}, 455 (1990)
  [Erratum-ibid.\ B {\bf 262}, 497 (1991)].

\bibitem{Paige:2003mg}
  F.~E.~Paige, S.~D.~Protopescu, H.~Baer and X.~Tata,
  arXiv:hep-ph/0312045.

\bibitem{Yamashita}
    S.~Yamashita, talk given at ``2005 International Linear Collider
    Physics and Deterctor Workshop and Second ILC Accelerator
    Worksnop'' (2005, Snowmass, Colorado), {\tt
    http://alcpg2005.colorado.edu}.

\bibitem{Boos:2004kh}
  E.~Boos {\it et al.}  [CompHEP Collaboration],
  Nucl.\ Instrum.\ Meth.\ A {\bf 534}, 250 (2004).

\bibitem{Gondolo:2004sc}
    P.~Gondolo {\it et al.},
    JCAP {\bf 0407}, 008 (2004).

\end{thebibliography}
\end{document}